\documentclass[11pt,a4paper]{article}
\pdfoutput=1
\usepackage{jheppub}

\usepackage{amsfonts}
\usepackage{amscd}
\usepackage{amssymb}
\usepackage{amsmath,bbm}
\usepackage{graphicx}
\usepackage{epsfig}
\usepackage{tabu}
\usepackage{latexsym}
\usepackage{mathtools}
\usepackage{hyperref}
\usepackage{tikz-cd}
\usepackage[vcentermath]{youngtab}
    
\def \be  {\begin{equation}}
\def \ee  {\end{equation}}
\def \bea {\begin{equation}\begin{aligned}}
\def \eea {\end{aligned}\end{equation}}
\def \ba  {\begin{eqnarray}}
\def \ea  {\end{eqnarray}}
\def \bb  {\begin{thebibliography}}
\def \eb  {\end{thebibliography}}
\def \lab #1 {\label{#1}}

\newcommand\ep{\epsilon}

\newcommand\sym{\mathrm{Sym}}

\newcommand\cA{\mathcal{A}}

\newcommand\cF{\mathcal{F}}
\newcommand\cG{\mathcal{G}}
\newcommand\cH{\mathcal{H}}
\newcommand\cI{\mathcal{I}}

\newcommand\cK{\mathcal{K}}
\newcommand\cL{\mathcal{L}}
\newcommand\cM{\mathcal{M}}
\newcommand\cN{\mathcal{N}}
\newcommand\cO{\mathcal{O}}

\newcommand\cV{\mathcal{V}}

\newcommand\al{\alpha}

\newcommand\tf{\mathfrak{t}}

\newcommand\C{\mathbb{C}}

\newcommand\bC{\mathbb{C }}

\newcommand\R{\mathbb{R}}

\newcommand\fm{\mathfrak{m}}
\newcommand\ft{\mathfrak{t}}

\newcommand\la{\langle}
\newcommand\ra{\rangle}

\newcommand\tr{\mathrm{Tr}}

\definecolor{cardinal}{rgb}{0.6,0,0}
\definecolor{darkgreen}{rgb}{0,0.5,0}
\definecolor{golden}{rgb}{0.92, 0.7, 0}
\definecolor{midnight}{rgb}{0, 0, 0.5}
\definecolor{darkblue}{rgb}{0.2, 0, 0.8}

\title{Twisted Hilbert Spaces of 3d Supersymmetric Gauge Theories}
\author[1,2]{Mathew Bullimore}
\author[2]{Andrea Ferrari}
\affiliation[1]{Mathematics Department, Durham University, Science Laboratories, \\ South Road, Durham,  
DH1 3LE, UK}
\affiliation[2]{Mathematical Institute, University of Oxford, Andrew Wiles Building,\\
Radcliffe Observatory Quarter,  Woodstock Road, Oxford, OX2 6GG, UK}
\abstract{We study aspects of 3d $\cN=2$ supersymmetric gauge theories on the product of a line and a Riemann surface. Performing a topological twist along the Riemann surface leads to an effective supersymmetric quantum mechanics on the line. We propose a construction of the space of supersymmetric ground states as a graded vector space in terms of a certain cohomology of generalized vortex moduli spaces on the Riemann surface. This exhibits a rich dependence on deformation parameters compatible with the topological twist, including superpotentials, real mass parameters, and background vector bundles associated to flavour symmetries. By matching spaces of supersymmetric ground states, we perform new checks of 3d abelian mirror symmetry.}

\begin{document}
\today
\maketitle


\section{Introduction}

\begin{figure}[htp]
\centering
\includegraphics[height=2cm]{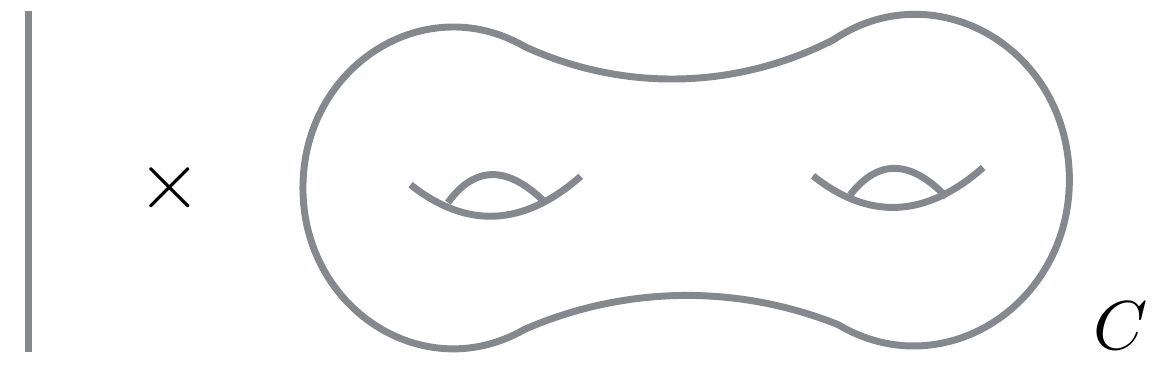}
\caption{We consider 3d $\cN=2$ supersymmetric gauge theories on the product of a line and a Riemann surface $C$.}
\label{fig:intro}
\end{figure}

We study aspects of 3d $\cN=2$ supersymmetric gauge theories on the product of a line and a Riemann surface $C$. We introduce a topological twist on the Riemann surface $C$, which preserves a pair of supercharges generating a supersymmetric quantum mechanics on the line with supermultiplets of the type obtained by the dimensional reduction of 2d $\cN=(0,2)$ supermultiplets. The setup is illustrated in figure~\ref{fig:intro}.

Following the philosophy of~\cite{Bullimore:2016hdc}, our ultimate goal is to find an effective supersymmetric quantum mechanics that captures correlation functions of the 3d $\cN=2$ supersymmetric gauge theory compatible with the topological twist on $C$. As a first step, in this paper we consider the space of supersymmetric ground states of the supersymmetric quantum mechanics, which we call the \emph{twisted Hilbert space}.

The supersymmetric ground states are charged under Fermion number and flavour symmetry. The twisted Hilbert space then transforms as $\mathbb{Z}_2$-graded or virtual representation of the flavour symmetry. Furthermore, it exhibits a rich dependence on supersymmetric deformation parameters obtained by coupling to background vectormultiplets for flavour symmetry in a manner compatible with the topological twist on $C$. In particular, for the flavour symmetry $G_f$ acting on chiral multiplets, we may turn on
\begin{enumerate}
\item A background $G_f$-vector bundle $E_f$ on $C$.
\item Real mass parameters $m_f$ valued in the Cartan subalgebra of $G_f$.
\end{enumerate}
In supersymmetric gauge theories with abelian factors in the gauge group, we can also introduce deformation parameters associated to topological flavour symmetries. In particular, a real mass for a $U(1)$ topological symmetry corresponds to a real FI parameter.

In this paper, we will focus on supersymmetric gauge theories with $G = U(1)$. In this case, the supersymmetric twisted Hilbert space decomposes as a direct sum of contributions labelled by the quantized flux $\fm\in \mathbb{Z}$ on $C$,
\be
\cH = \bigoplus\limits_{\fm \in \mathbb{Z}} \cH_{\fm} \, .
\ee
Supersymmetric ground states in $\cH_\fm$ have charge $\fm$ under the topological flavour symmetry. Each summand is a finite-dimensional $\mathbb{Z}_2$-graded representation of the remaining flavour symmetry $G_f$ acting on the chiral multiplets.
It is captured by an effective supersymmetric quantum mechanics whose target space is the moduli space $\cM_\fm$ of solutions to a set of generalized vortex equations on $C$ with flux $\fm$ in the presence of the background vector bundle $E_f$. 

The generalized vortex moduli spaces $\cM_\fm$ are K\"ahler and may or may not be compact, depending on the matter content and flux $\fm \in \mathbb{Z}$. However, provided the moduli space $\cM_\fm$ is smooth, we propose that the space of supersymmetric ground states can be understood in terms of an $L^2$-cohomology
\be
\cH_\fm = H_{\bar\partial_{m_f}+ \delta}^{0,\bullet} \, (\cM_\fm , \cF_\fm) \, ,
\ee
where $\cF_\fm$ denotes a $\mathbb{Z}_2$-graded vector bundle that receives contributions from supersymmetric Chern-Simons terms, a background line bundle for the topological flavour symmetry on $C$, and the quantization of Fermion zero modes. The differential is a sum of a conjugated Dolbeault operator,
\be
\bar\partial_{m_f} = e^{-h_f} \cdot \bar\partial \cdot e^{h_f} \, ,
\ee
where the real superpotential $h_f = m_f \cdot \mu_f$ is constructed from the moment map $\mu_f$ for the action of the flavour symmetry $G_f$ on $\cM_\fm$, and an additional contribution $\delta$ from a 3d superpotential. This $L^2$-cohomology can depend intricately on the choice of background vector bundle $E_f$ and real mass parameters $m_f$.

An important consistency check for our proposal is to reproduce the supersymmetric twisted index on $S^1 \times C$, which may be computed in the supersymmetric gauge theory using localization~\cite{Benini:2015noa,Benini:2016hjo,Closset:2016arn}. This should be recovered from the supersymmetric twisted Hilbert space by computing the graded trace,
\bea
\cI & = \mathrm{Tr}_{\cH}(-1)^F q^{J_T}x^{J_f} \\
& = \sum_{\fm \in \mathbb{Z}}  q^\fm \, \mathrm{Tr}_{\cH_\fm}(-1)^F x^{J_f} \, ,
\label{eq:intro-index}
\eea
where $J_T$, $J_f$ are the generators of the topological and remaining $G_f$ flavour symmetry respectively, and the parameters $q$ and $x$ are the respective fugacities. Our proposal therefore provides a new interpretation of the supersymmetric twisted index as a generating function of equivariant Euler characters associated to the generalized vortex moduli spaces $\cM_\fm$.

We emphasize, however, that the supersymmetric twisted Hilbert space exhibits information and structure that goes beyond the supersymmetric twisted index:
\begin{itemize}
\item There can be dramatic cancellations in computing the supersymmetric index via~\eqref{eq:intro-index}, particularly on Riemann surfaces of genus $g>0$. 
\item The supersymmetric twisted Hilbert space is sensitive to 3d superpotential deformations via the differential $\delta$, which removes pairs of supersymmetric ground states whose contribution to the supersymmetric twisted index cancel out.
\item The supersymmetric twisted Hilbert space may jump across hyperplanes in the space of real mass parameters $m_f$ where there are non-compact massless degrees of freedom. On the other hand, the supersymmetric index is a meromorphic function with poles on these hyperplanes. The same remark applies to real FI parameters for topological flavour symmetries.
\item The supersymmetric twisted Hilbert space depends on a choice of holomorphic vector bundle $E_f$ on $C$ for the flavour symmetry $G_f$, while the supersymmetric twisted index depends only on its Chern class. The same remark applies to background line bundles associated to topological flavour symmetries.
\end{itemize}
The supersymmetric twisted Hilbert space can therefore provide a more refined check of supersymmetric dualities such as 3d mirror symmetry~\cite{Aharony:1997bx,deBoer:1997kr,deBoer:1996ck,Intriligator:1996ex}.

In order to illustrate some of these points, we provide a brief appetizer. Let us consider a supersymmetric $U(1)$ Chern-Simons theory at level $+\frac{1}{2}$ with a chiral multiplet of charge $+1$. In section~\ref{sec:U(1)1/2}, we will show that the supersymmetric twisted Hilbert space with real FI parameter $\zeta >0$ at infinite coupling is given by
\be
\cH  = \bigoplus_{\fm = 1-g}^\infty q^\fm \bigoplus_{j=0}^{\fm+g-1} \wedge^{j}(\C^g) \, ,
\label{eq:intro-u(1)}
\ee
where $g>0$ is the genus of $C$ and the parameter $q$ keeps track of the grading by the topological flavour symmetry. Notice that there are non-vanishing contributions from an infinite number of fluxes, $\fm \geq 1-g$. On the other hand, the supersymmetric twisted index is a finite Laurent polynomial
\bea
\cI = q^{1-g}(1-q)^{g-1} \, ,
\eea
with the contributions from fluxes $\fm \geq 0$ cancelling out in the trace. Nevertheless, in section~\ref{sec:U(1)-chiral-mirror} we will demonstrate that equation~\eqref{eq:intro-u(1)} agrees with the supersymmetric twisted Hilbert space of a single chiral multiplet with a positive real mass parameter. Furthermore, we extend this agreement to include a holomorphic line bundle for the flavour symmetry. This constitutes a new check of the simplest 3d mirror symmetry.

Finally, let us mention some connections to related research. First, an important motivation for this work is to generate new realisations of the correspondence between 3d $\cN=2$ theories and three-manifolds~\cite{Dimofte:2011py,Dimofte:2011ju,Cecotti:2011iy} and we note that computations of twisted Hilbert spaces have appeared in this context in~\cite{Gukov:2017kmk,Gukov:2016gkn}. Secondly, it would be interesting to consider potential applications to supersymmetric quantum Hall systems on a Riemann surface~\cite{Iengo:1993cs,Klevtsov:2017aa,Tong:2003vy,Tong:2015xaa}.

The paper is structured as follows. In section~\ref{sec:susy-qm}, we review aspects of $\cN=(0,2)$ type supersymmetric quantum mechanics, emphasising how the space of supersymmetric ground states depends on various types of deformation parameters. In section~\ref{sec:3d-setup}, we consider the topological twist of 3d $\cN=2$ theories on a Riemann surface $C$ and explain the general structure of the effective supersymmetric quantum mechanics. In section~\ref{sec:examples}, we compute the supersymmetric twisted Hilbert space in a variety of examples, including theories of chiral multiplets with superpotentials and abelian supersymmetric gauge theories. In section~\ref{sec:mirror}, we perform new checks of 3d mirror symmetry using the results from section~\ref{sec:examples}. Finally, in section~\ref{sec:disc} we conclude with a discussion of directions for further research.


\section{Supersymmetric Quantum Mechanics}
\label{sec:susy-qm}

We review supersymmetric quantum mechanics with supermultiplets that arise from the dimensional reduction of $\cN=(0,2)$ supermultiplets in two dimensions, emphasizing those aspects that will be important in applications to 3d $\cN=2$ theories on a Riemann surface. For further background and examples of this class of supersymmetric quantum mechanics we refer the reader to~\cite{Hori:2014tda,Wong:2015qnf}.


\subsection{Setup}
\label{sec:qm-setup}

A supersymmetric quantum mechanics has odd generators $Q$ and $\bar Q$ that are adjoint with respect to a hermitian inner product on the Hilbert space. We suppose the supersymmetric quantum mechanics has flavour symmetry $G_f$ with conserved charges $J_f \in \mathfrak{t}_f^*$ and introduce an associated real mass parameter $m_f \in \mathfrak{t}_f$. Here $\tf_f$ is the Cartan subalgebra of $G_f$. The supersymmetry algebra is then
\bea
\{ Q, \bar Q \}  & = 0 \\
\{ Q , \bar Q \}  & = H  -  m_f \cdot J_f \\
 \{ \bar Q,  \bar Q \}  & = 0 \, ,
\label{eq:susy-hilb}
\eea
where $H$ is the hamiltonian. 

We define supersymmetric ground states as those annihilated by $H-m_f\cdot J_f$. A standard argument shows that the spectrum of $H - m_f \cdot J_f$ is non-negative and that supersymmetric ground states are equivalently annihilated by both of the supercharges. We assume that the spectrum is gapped, in which case the supersymmetric ground states have another equivalent description as the cohomology of either supercharge $Q$ or $\bar Q$. 

The requirement that the spectrum is gapped may place constraints on the mass parameters $m_f$. We will denote the subspace of mass parameters where the spectrum is gapped by $\mathfrak{c}_f \subset \mathfrak{t}_f$. If $\mathfrak{c}_f \neq \mathfrak{t}_f$, this subspace typically consists of a union of chambers $\mathfrak{c}_f = \bigcup_\al \mathfrak{c}_\al \subset \ft_f$ cut out by hyperplanes where there are non-compact massless degrees of freedom.

In this paper, we will compute the space of supersymmetric ground states $\cH$ as the cohomology of the supercharge $\bar Q$.
Since $(-1)^F$ and $J_f$ commute with the supercharges, the space of supersymmetric ground states is graded by Fermion number and flavour symmetry. Alternatively, we can say that it is a $\mathbb{Z}_2$-graded or virtual representation of the flavour symmetry $G_f$.

It will be important to understand how the supersymmetric ground states change as the real mass parameters $m_f$ are varied. In the examples encountered below, the supercharges obey
\bea
\partial_{m_f} Q & = + [ \mu_f , Q ] \\
\partial_{m_f} \bar Q & = - [ \mu_f , \bar Q ] \, ,
\eea
where $\mu_f \in \mathfrak{t}_f^*$ is a hermitian operator. This is an $A$-type deformation in the notation of~\cite{Gaiotto:2016hvd}. In particular, the operator $\partial_{m_f} + \mu_f$ commutes with $\bar Q$ and descends to a complex flat Berry connection on the sheaf of supersymmetric ground states over $\mathfrak{c}_f \subset \tf_f$. Put simply, while the wavefunctions of the supersymmetric ground states will depend explicitly on the real mass parameters, $\cH$ remains constant as a graded vector space provided the spectrum remains gapped. Therefore, we associate a space of supersymmetric vacua $\cH_\al$ to each chamber $\mathfrak{c}_\al$.

We will also encounter examples of $B$-type deformations of the supersymmetric quantum mechanics~\cite{Gaiotto:2016hvd}, where the supercharges depend holomorphically or anti-holomorphically on a set complex parameters $u$,
\bea
\partial_{u} Q & = 0 \\
\partial_{\bar u} \bar Q & = 0 \, .
\eea
In particular, the derivative $\partial_{\bar u}$ commutes with $\bar Q$ and provided the system is gapped will descend to a holomorphic Berry connection on the sheaf of supersymmetric ground states over the complex space parametrized by $u$.

The flavoured supersymmetric index is defined as a graded trace over the full Hilbert space of the supersymmetric quantum mechanics,
\bea
\cI = \tr (-1)^F  e^{-2\pi\beta H} e^{-2\pi i \beta A_f \cdot J_f} \, .
\eea
In the euclidean path integral construction of the supersymmetric index, $\beta$ is the radius of the circle and $e^{-2\pi i \beta A_f \cdot J_f}$ is a background Wilson line for the flavour symmetry. A standard argument shows that only supersymmetric ground states contribute to the supersymmetric index and therefore in each chamber $\mathfrak{c}_\al$ we obtain an expression
\bea
\cI_\al & = \tr_{\cH_\al} (-1)^F e^{-2\pi \beta (m_f + i A_f ) \cdot J_f} \\
  & = \tr_{\cH_\al} (-1)^F x^{J_f} \, ,
  \label{eq:index-trace}
\eea
where 
\be
x =e^{- 2\pi \beta( m_f + i A_f)} 
\label{eq:exp}
\ee
is valued in the complexified maximal torus of the flavour symmetry $G_f$. The supersymmetric index can therefore be expressed as a graded trace over $\cH_\al$.

The supersymmetric index $\cI_\al$ computed in equation~\eqref{eq:index-trace} will in general yield a different Laurent polynomial in $x$ in each chamber $\mathfrak{c}_\al$. However, they correspond to Laurent expansions of the same meromorphic function $\cI(x)$ in the different chambers $\mathfrak{c}_\al$ under the identification~\eqref{eq:exp}. This meromorphic function then has poles on the hyperplanes separating these chambers. In the case $\mathfrak{c}_f = \mathfrak{t}_f$, the supersymmetric index is a finite Laurent polynomial in $x$.

Finally, the supersymmetric index is insensitive to $B$-type deformations.


\subsection{Geometric Model}
\label{sec:geometric-model}

We now consider a general class of supersymmetric quantum mechanics of the above type that arise from supersymmetric sigma models. The construction of these supersymmetric quantum mechanics has much in common with the construction of 2d $\cN=(0,2)$ supersymmetric sigma models~\cite{Witten:1993yc}.

We consider a supersymmetric sigma model specified by the following data:
\begin{itemize}
\item A complex manifold $M$ with hermitian metric.
\item A $\mathbb{Z}$-graded hermitian vector bundle $F$.
\item A holomorphic differential $\delta : F\to F$ of degree $+1$ obeying $\delta^2 = 0$.
\end{itemize}
The full Hilbert space of the supersymmetric quantum mechanics consists of smooth square-integrable sections of
\be
\Omega^{0,\bullet}(M) \, \otimes \, F 
\label{eq:sections}
\ee
with respect to the hermitian inner product
\be
\la \al , \beta \ra = \int_M \bar \al \wedge * \beta \, .
\label{eq:inner-prod}
\ee
Here, $*$ denotes the Hodge star operator on $M$ and contraction along fiber directions using the hermitian metric on $F$ is understood. 

The supersymmetric quantum mechanics has an $R$-symmetry transforming the supercharges $Q$, $\bar Q$ with charge $-1$, $+1$ respectively. Referring to the above geometric data, this R-symmetry can be identified with the sum of the form degree on the target space $M$ and the $\mathbb{Z}$-grading on the hermitian vector bundle $F$, modulo an additive constant. 

In this paper, we will only keep track of the Fermion number $(-1)^F$. In particular, we regard $F$ as a $\mathbb{Z}_2$ graded vector bundle with decomposition $F = F_e \oplus F_o$ into even and odd components. The Fermion number $(-1)^F$ in the supersymmetric quantum mechanics is then given the sum of the form degree and the $\mathbb{Z}_2$-grading on $F$:
\begin{itemize}
\item $(-1)^q$ for a section of $\Omega^{0,q}(M) \otimes F_e$ and 
\item $(-1)^{q+1}$ for a section of $\Omega^{0,q}(M) \otimes F_o$ \, .
\end{itemize}

Let $G_f$ denote the group of isometries of $M$ that lift to an equivariant action on $F$ preserving its hermitian metric and commuting with the holomorphic differential $\delta$. This is the flavour symmetry of the supersymmetric quantum mechanics. At this point, we assume that $M$ is K\"ahler and there exists a corresponding real moment map $\mu_f \in \mathfrak{t}_f^*$. We may then introduce an $A$-type deformation of the supersymmetric quantum mechanics by real mass parameters $m_f \in \tf_f$, which can be understood as a real superpotential
\be
h_f = m_f \cdot \mu_f \, .
\ee
This superpotential is the moment map for the $U(1)_{m_f} \subset G_f$ isometry generated by the mass parameters $m_f$~\cite{witten-hol-morse}. 

Let $\mathfrak{c}_f \subset \mathfrak{t}_f$ denote the mass parameters where the fixed locus of the $U(1)_{m_f}$ isometry of $M$ is compact and the spectrum of the supersymmetric quantum mechanics is gapped. If $M$ is non-compact, this is a disjoint union of chambers $\mathfrak{c}_f = \bigcup_\al \mathfrak{c}_\al$
cut out by hyperplanes. If $M$ is compact, $\mathfrak{c}_f = \mathfrak{t}_f$.

The supercharges are identified with
\bea
Q  & = \bar\partial^\dagger_{m_f} + \delta^\dagger \\
\bar Q  & = \bar\partial_{m_f} + \delta \; .
\label{eq:Q-def}
\eea
where 
\bea
\bar\partial^\dagger_{m_f} & := e^{h_f} \cdot \bar \partial^\dagger \cdot e^{-h_f} \\
\bar\partial_{m_f} & := e^{-h_f} \cdot \bar \partial \cdot e^{h_f}
\label{eq:conj-dolbeault}
\eea
and $\bar\partial,\bar\partial^\dagger$ denote respectively the twisted Dolbeault operator acting on sections of~\eqref{eq:sections} and its adjoint with respect to the hermitian inner product~\eqref{eq:inner-prod}.
Finally, $\delta^\dagger$ is the adjoint of holomorphic differential $\delta$ with respect to the hermitian metric on $F$. Note that the supercharge $\bar Q$ depends holomorphically on deformations of the holomorphic vector bundle $F$ and the differential $\delta$ : they are $B$-type deformations of the supersymmetric quantum mechanics. 

These supercharges obey the supersymmetry algebra~\eqref{eq:susy-hilb} where $H$ is a deformation of the Dolbeault Laplacian and $m_f \cdot J_f$ is the Lie derivative along the vector field generating $U(1)_{m_f}$.

Provided the mass parameters lie in $\mathfrak{c}_f \subset \tf_f$ and the spectrum of the supersymmetric quantum mechanics is gapped, the space of supersymmetric ground states can be identified with $L^2$-cohomology of the supercharge $\bar Q$, which we write schematically as
\be
H^{0,\bullet}_{\bar\partial_{m_f} + \delta}(M,F) \, .
\ee
Due to the exponential dependence of the supercharge on the mass parameter $m_f$ and the condition of square-normalizability, the computation of this cohomology for a non-compact target space $M$ will generally yield a different space of supersymmetric ground states $\cH_\al$ in each chamber $\mathfrak{c}_\al$. In other words, the space of supersymmetric ground states may jump across hyperplanes in $\tf_f$ where there are massless non-compact degrees of freedom.

However, if $M$ is compact, the spectrum of the supersymmetric quantum mechanics is gapped for any $m_f$ and the space of supersymmetric vacua is constant on $\mathfrak{t}_f$. In this case, we can set $m_f = 0$ and identify the space of supersymmetric ground states with the regular hypercohomology,
\be
\cH = H^{0,\bullet}_{\bar\partial + \delta}(M,F)  \, .
\ee 

Let us finally consider the supersymmetric index in this class of supersymmetric quantum mechanics. The supersymmetric index $\cI_\al$ in each chamber computes the equivariant character of $\cH_\al$ as a virtual representation of the flavour symmetry $G_f$. This index is independent of the differential and can be identified with an equivariant Euler character for $L^2$-cohomology classes of the conjugated Dolbeault operator $\bar \partial_{m_f}$ in equation~\eqref{eq:conj-dolbeault}. If $M$ is compact, the supersymmetric index $\cI$ coincides with the regular equivariant Euler character $\chi(M,F)$.


\subsection{Examples}
\label{sec:qm-examples}

\subsubsection{Chiral Multiplets}
\label{sec:qm-chirals}

Our first example is a single chiral multiplet $(\phi,\psi)$ with real a mass parameter $m_f$ for the $U(1)_f$ flavour symmetry. This model is a supersymmetric complex harmonic oscillator. In canonical quantization, the complex Fermion obeys $\{\psi,\bar\psi\} = 1$ and the supercharges take the form
\bea
Q & = \psi \left( -\frac{\partial}{\partial \phi} + m_f\bar\phi \right)  \\
\bar Q & =   \bar\psi \left( + \frac{\partial}{\partial \bar\phi} + m_f\phi \right) \, .
\eea
The supercharges obey~\eqref{eq:susy-hilb} with
\bea
H & = - \frac{\partial^2}{\partial\phi \partial \bar\phi} + m_f^2 | \phi|^2 - \frac{1}{2} m_f [\psi ,\bar\psi]  + m_f\kappa \\
J_f & = \phi \frac{\partial}{\partial \phi} -  \bar\phi \frac{\partial}{\partial \bar\phi}+ \frac{1}{2}[\psi ,\bar\psi] + \kappa \, .
\label{eq:HandJ1}
\eea
Note that while the supercharges and the combination $H-m_fJ_f$ are unambiguous, $H$ and $J_f$ individually depend on a normal ordering constant $\kappa$, which can be understood as a supersymmetric Chern-Simons term for the $U(1)_f$ flavour symmetry~\cite{Assel:2015nca,Elitzur:1985xj}.

\begin{figure}[htp]
\centering
\includegraphics[height=1.5cm]{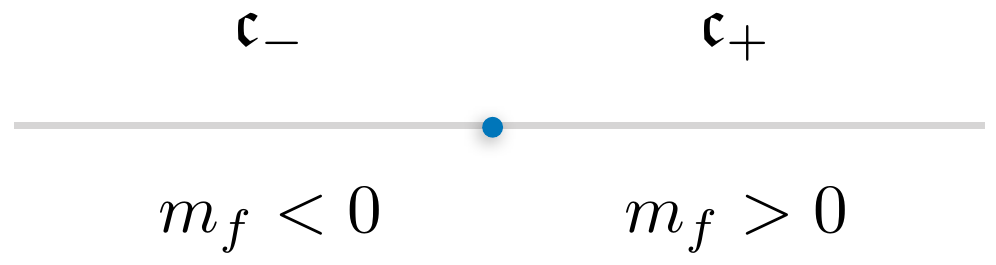}
\caption{Chambers $\mathfrak{c}_+ = \{ m_f>0\}$ and $\mathfrak{c}_- = \{ m_f<0\}$ for a single chiral multiplet.}
\label{fig:chambers1}
\end{figure}

The supersymmetric ground states wavefunctions are annihilated by both supercharges. The gapped region of parameter space consists of two chambers $\mathfrak{c}_\pm \subset \mathfrak{t}_f = \mathbb{R}$ corresponding to $m_f>0$ and $m_f<0$ respectively - see figure~\ref{fig:chambers1}. Choosing the Fock vacuum annihilated by $\psi$, the normalizable ground state wavefunctions are
\bea
\mathfrak{c}_+ & \quad : \quad e^{-h_f}\phi^n  \\
\mathfrak{c}_- & \quad : \quad e^{h_f}\bar\phi^n \bar \psi \, ,
\label{eq:chiral-vac-wav}
\eea
where $n\geq 0$. Here we have defined a superpotential
\be
h_f = m_f |\phi|^2 \, .
\ee

The supersymmetric ground state wavefunctions can also be viewed as harmonic representatives of $L^2$-cohomology classes for the supercharge $\bar Q$. We denote the associated cohomology classes by $[\, \phi^n\, ]$ in the chamber $\mathfrak{c}_+$ and $[\, \bar\phi^n \, ]$ in the chamber $\mathfrak{c}_-$. In the second chamber, it is important to remember the presence of the Fermion $\bar\psi$, which is suppressed in our notation. Since the operator $\phi$ commutes with the supercharge $\bar Q$, it has a well-defined action on these cohomology classes, 
\bea
& \mathfrak{c}_+ \quad : \quad \phi \cdot[\, \phi^n\, ] = [\, \phi^{n+1}\, ]  \\
& \mathfrak{c}_- \quad : \quad \phi \cdot[\, \bar\phi^n\, ] = [\, \bar\phi^{n-1}\, ] \, ,
\eea
which is compatible with the $U(1)_f$ flavour symmetry.

Although the supersymmetric ground state wavefunctions depend on $m_f$, as a vector space graded by $(-1)^F$ and $U(1)_f$, the space of supersymmetric ground states is constant in each chamber,
\bea
\cH_{+} \; & = \; x^{\kappa + \frac{1}{2}}   \displaystyle\bigoplus\limits_{j = 0}^\infty x^j \C \\
\cH_{-} \; & = \; - x^{\kappa-\frac{1}{2}} \displaystyle\bigoplus\limits_{j = 0}^\infty x^{-j} \C  \,.
\eea
Here we have introduced a formal parameter $x \in \C^*$ to keep track of the $U(1)_f$ charge measured by $J_f$. Note that there is a choice of Fermion number for the Fock vacuum, and we have assigned Fermion number zero to the Fock vacuum annihilated by $\psi$. 

The supersymmetric index computed in each chamber is
\bea
\cI_+ \; & = \; x^{\kappa + \frac{1}{2}} \displaystyle\sum\limits_{j=0}^\infty x^j \\
\cI_- \; & = \; - x^{\kappa-\frac{1}{2}} \displaystyle\sum\limits_{j=0}^\infty x^{-j} \, .
\eea
Recalling that in computing the supersymmetric index we identify $x = e^{-2\pi \beta(m_f + i A_f )}$, this corresponds to the expanding the same meromorphic function
\be
\cI(x) = \frac{x^{\kappa + \frac{1}{2}}}{1-x}
\label{eq:1chiral-index}
\ee
in the appropriate regime, namely $|x|<1$ in $\mathfrak{c}_+$ and $|x|>1$ in $\mathfrak{c}_-$. This expression coincides with the 1-loop determinant for a chiral multiplet on a circle with a background supersymmetric Chern-Simons term for $U(1)_f$ at level $\kappa$. 

This example can be extended to $N$ chiral multiplets $(\phi_j,\psi_j)$ with flavour symmetry $G_f = U(N)$ and mass parameters $m_f = (m_1,\ldots,m_N) \in \mathfrak{t}_f$. There are now massless degrees of freedom on all coordinate hyperplanes in $\mathfrak{t}_f = \mathbb{R}^N$. Removing these hyperplanes, the gapped region $\mathfrak{c}_f = \bigcup_\al \mathfrak{c}_\al$ consists of $2^N$ disjoint chambers 
\be
\mathfrak{c}_\alpha = \begin{cases} m_j > 0 & \quad \alpha_j = + \\
m_j<0 & \quad \alpha_j = - \end{cases} \, ,
\label{eq:chiral-gapped}
\ee
labelled by a sign vector $\alpha = (\alpha_1,\ldots,\alpha_N)$. 

The space of supersymmetric grounds states in each chamber is
\be
\cH_\alpha = \bigotimes_{j=1}^N \cH_{\alpha_j}
\ee
where
\bea
\cH_{\alpha_j} \; = \;
\begin{cases}
\; x_j^{\kappa + \frac{1}{2}}   \displaystyle\bigoplus\limits_{n = 0}^\infty x_j^n \C \quad &  \alpha_j = + \\
\; - x_j^{\kappa-\frac{1}{2}} \displaystyle\bigoplus\limits_{n = 0}^\infty x_j^{-n} \C \quad & \alpha_j = - \,.
\end{cases}  
\eea
We have chosen the same normal ordering constant $\kappa$ for each chiral multiplet to preserve the underlying $G_f = U(N)$ flavour symmetry. As expected, the result reproduces the expansion of the supersymmetric index
\be
\prod_{j=1}^N \frac{x_j^{\kappa+\frac{1}{2}}}{1-x_j}
\ee
in the appropriate regime $|x_j|^{\alpha_j} <1$.

This model can be understood as a supersymmetric sigma model to $M = \C^N$ with the standard flat K\"ahler metric, supplemented by a hermitian line bundle $F = K_{\mathbb{C}^N}^{\kappa+1/ 2}$. The flavour symmetry $G_f = U(N)$ corresponds to the isometries of $\C^N$. Introducing real mass parameters corresponds to a superpotential 
\be
h_f = \sum_j m_j |\phi_j|^2\, ,
\ee
which is the moment map 
for the $U(1)_{m_f}$ isometry generated by $m_f \in \tf_f$. The chambers $\mathfrak{c}_\al$ correspond to values of the mass parameters where the fixed locus of $U(1)_{m_f}$ is compact, namely the origin of $\mathbb{C}^N$.


\subsubsection{Fermi Multiplets}
\label{sec:qm-Fermi}

Let us now consider a single Fermi multiplet $(\eta,F)$ with a real mass parameter $m_f$ for the $U(1)_f$ flavour symmetry. In canonical quantization, the complex Fermion obeys $\{ \eta , \bar\eta \} = 1$ and
\bea
H & =\frac{m_f}{2}[\eta,\bar\eta] + m_f \kappa   \\
J_f & = \frac{1}{2}[\eta,\bar\eta] + \kappa \, .
\label{eq:HJ-Fermi}
\eea
The combination $H - m_f J_f = 0$ is again unambiguous, whereas $H$ and $J_f$ individually depend on a normal ordering constant $\kappa$. 

In canonical quantization, we can choose a Fock vacuum or reference state $|0\ra$ annihilated by the Fermion $\eta$ and assign it Fermion number $0$. The supersymmetric ground states are then $|0\ra$ and $\bar\eta|0\ra$ with flavour charge $\kappa+\frac{1}{2}$ and $\kappa-\frac{1}{2}$ respectively, as measured by $J_f$. We therefore find
\be
\cH = x^{\kappa+1/2} \C  - x^{\kappa - 1/2} \C \, ,  
\ee
in agreement with the supersymmetric index
\be
\cI = x^{\kappa+1/2} - x^{\kappa-1/2} \, .
\label{eq:1Fermi-index}
\ee 

In the quantization of Fermi multiplets, there is a notational freedom to choose the Fock vacuum or reference state $|0\ra$ to be annihilated by $\eta$ or $\bar\eta$. In more complicated examples below, we will use this freedom to choose the representation that is most convenient for enumerating the supersymmetric ground states.  The Fermion number assigned to this Fock vacuum is, however, meaningful and sets the Fermion number grading of supersymmetric ground states. This corresponds to an overall sign in the supersymmetric index. The reader is forewarned that we will typically omit the reference state $|0\ra$ from our notation.

\subsubsection{Superpotentials}

We now present a number of examples that couple chiral and Fermi multiplets with holomorphic superpotentials and will reappear in computations relevant for 3d $\cN=2$ theories in section~\ref{sec:examples}. 

Let us first consider a chiral multiplet $(\phi,\psi)$ coupled to a Fermi multiplet $(\eta,F)$ with the following $J$-term superpotential
\be
J(\phi) = u \phi \, ,
\label{eq:j-linear}
\ee 
where the complex mass parameter $u$ can be regarded as a vacuum expectation value for a background chiral multiplet. The model preserves a $G_f = U(1)$ flavour symmetry under which $(\phi,\psi)$ have charge $+1$ and $(\eta,F)$ have charge $-1$. We introduce a corresponding real mass parameter $m_f \neq 0$. 

In canonical quantization, the supercharge
\be
\bar Q = \bar Q^{(0)} + \bar Q^{(1)}
\ee
is a sum of two contributions
\be
\bar Q^{(0)} = \left( \frac{\partial}{\partial \bar\phi} + m_f\phi \right) \bar\psi  \qquad \bar Q^{(1)}  = J(\phi)  \eta \, ,
\ee
where $\bar Q^{(0)}$ is the contribution from the chiral multiplet and $\bar Q^{(1)}$ is the additional contribution from the Fermi multiplet with $J$-term superpotential. Note that in this model there is both a real $A$-type parameter $m_f$ and a complex $B$-type parameter $u$.

We assign Fermion number zero to the Fock vacuum annihilated by $\psi$ and $\eta$. First, the cohomology of $\bar Q^{(0)}$ consists of the supersymmetric ground states of the chiral multiplet tensored with those of the Fermi multiplet,
\bea
\mathfrak{c}_+ & \quad : \quad  [\,\phi^n\,] \, , \, [\,\phi^n\,] \, \bar\eta \quad &&  n \geq 0 \\
\mathfrak{c}_- & \quad : \quad [\,\bar\phi^n\,] \, , \,  [\,\bar\phi^n\,]  \, \bar \eta \quad && n \geq 0 \, .
\label{eq:chiral-Fermi-step1}
\eea
If $u = 0$, the computation ends here and there is an infinite number of supersymmetric ground states.
Assuming $u \neq 0$, a short spectral sequence argument shows that the cohomology of the total supercharge $\bar Q$ is equivalent to the cohomology of $\bar Q^{(1)}$ acting on the states~\eqref{eq:chiral-Fermi-step1}. 
This is computed as follows:
\begin{itemize}
\item $\mathfrak{c}_+$: $\bar Q^{(1)}$ removes pairs $[\,\phi^{n+1}\,]$ and $[\,\phi^n\,] \, \bar\eta $ with $n\geq0$ leaving only $[\, 1\, ]$. 
\item $\mathfrak{c}_-$: $\bar Q^{(1)}$ removes pairs $[\,\bar\phi^{n}\,]$ and $[\,\bar\phi^{n+1}\,] \, \bar\eta$ with $n\geq 0$ leaving only $[\, 1\, ]\, \bar\eta$. 
\end{itemize}
We therefore find that for $u \neq 0$ there is a unique supersymmetric ground state and, setting normal ordering constants $\kappa=0$, $\cH_\alpha =\C$ in both chambers $\alpha = \pm$. 

Let us compare this result with the supersymmetric index. This is computed by multiplying the contributions from a chiral multiplet of charge $+1$ and a Fermi multiplet of charge $-1$, with the result $\cI = 1$ for both $u = 0$ and $u \neq 0$. In summary, the space of supersymmetric vacua is sensitive to the $J$-term superpotential whereas the supersymmetric index is not.

\begin{figure}[htp]
\centering
\includegraphics[height=3cm]{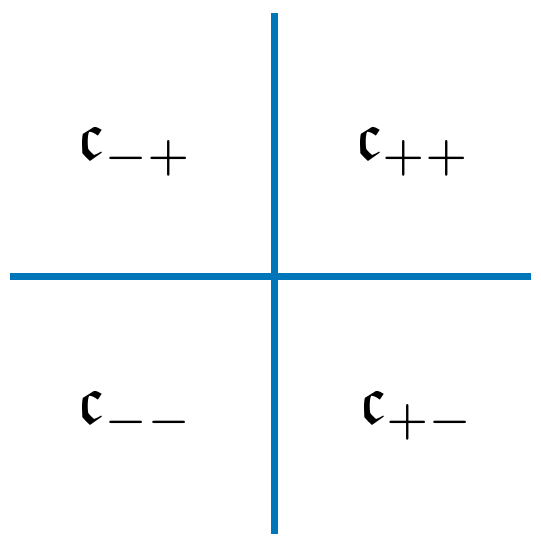}
\caption{Chambers in the space of real mass parameters $(m_1,m_2) \in \mathfrak{t}_f = \R^2$ for two chiral multiplets $\phi_1, \phi_2$ and Fermi multiplet with superpotential $J = \phi_1\phi_2$.}
\label{fig:chambers2}
\end{figure}

Let us now consider a second example with a pair of chiral multiplets $\phi_1$, $\phi_2$ coupled to a Fermi multiplet $\eta$ with superpotential $J(\phi) = \phi_1\phi_2$\footnote{As in our previous example, we could introduce a dimensionless complex parameter $u$ in the superpotential. Since the Hilbert space of supersymmetric vacua will not depend on this parameter provided $u\neq1$, we set $u = 1$ for convenience.}. This preserves a $U(1)_1\times U(1)_2$ flavour symmetry whose charges can be chosen as follows,
\begin{center}
\begin{tabular}{c|c|c}
& $U(1)_1$ & $U(1)_2$   \\
\hline
$\phi_1$ & $1$ & $0$  \\
$\phi_2$ & $0$ & $1$  \\
$\eta$ & $-1$ & $-1$  \\
\end{tabular}.
\end{center}
Introducing real mass parameters $m_1,m_2$ for the flavour symmetry, there are four chambers $\mathfrak{c}_{\alpha} \subset \mathfrak{t}_f = \R^2$ labelled by a sign vector $\alpha = (\alpha_1,\alpha_2)$ - see figure~\ref{fig:chambers2}. We again choose the Fock vacuum annihilated by $\psi_1,\psi_2$ and $\eta$.

As in our previous example, the supercharge is a sum of contributions from the chiral multiplets and the superpotential. The cohomology of $\bar Q^{(0)}$ is again the tensor product of supersymmetric ground states for the chiral and Fermi multiplets. Let us first consider the chamber $\mathfrak{c}_{++}$, in which we compute the cohomology of $\bar Q^{(1)} = \phi_1\phi_2 \eta $ acting on
\be
[\,\phi_1^{n_1}\phi_2^{n_2}\,] \, , \, [\,\phi_1^{n_1}\phi_2^{n_2}\,]\bar\eta
\ee
for $n_1,n_2 \geq 0$. The differential annihilates any $[\,\phi_1^{n_1}\phi_2^{n_2}\,]$ and sends the state $[\,\phi_1^{n_1}\phi_2^{n_2}\,]\bar\eta$ to $[\,\phi_1^{n_1+1}\phi_2^{n_2+1}\,]$. The remaining states in cohomology are
\bea
& \left[\,\phi_1^{n_1}\,\right]  \qquad n_1 \geq 0 \\
& \left[\,\phi_2^{n_2}\,\right]  \qquad n_2 >0 \, ,
\eea
and therefore
\be
\cH_{++} = \bigoplus\limits_{n_1\geq 0} x_1^{n_1} \C \oplus \bigoplus\limits_{n_2 > 0} x_2^{n_2} \C  \, ,
\ee
where again we set $\kappa=0$.

In the chamber $\mathfrak{c}_{+-}$, we compute the cohomology of $\bar Q^{(1)} = \phi_1\phi_2 \eta $ acting on
\be
[\,\phi_1^{n_1}\bar\phi_2^{n_2}\,] \, , \, [\,\phi_1^{n_1}\bar\phi_2^{n_2}\,]\bar\eta
\ee
with $n_1,n_2\geq 0$. The differential annihilates any $[\,\phi_1^{n_1}\bar\phi_2^{n_2}\,]$ and sends the state $[\,\phi_1^{n_1}\bar\phi_2^{n_2}\,]\bar\eta$ to $[\,\phi_1^{n_1+1}\bar\phi_2^{n_2-1}\,]$ if it exists. The remaining states in cohomology are
\bea
&\left[\,\phi_1^{n_1}\,\right] \bar\eta \quad && : \quad n_1 \geq 0 \\
&\left[\,\bar\phi_2^{n_2}\,\right] \quad && : \quad n_2 \geq 0
\eea
and therefore
\be
\cH_{+-} = \bigoplus\limits_{n_1 \geq 0} x_1^{n_1+1} \C \oplus \bigoplus\limits_{n_2 \geq  0} (-x_2^{-n_2-1}) \C \, .
\ee
There are similar results on regions $\mathfrak{c}_{-+}$ and $\mathfrak{c}_{--}$.

This is consistent with expanding the supersymmetric index
\be
\cI = \frac{1-x_1x_2}{(1-x_1)(1-x_2)}
\ee
in the appropriate chambers. The supersymmetric index does not detect the presence of the $J$-term superpotential, except through the determination of the flavour symmetry and associated mass parameters. In particular, the contributions from states removed in pairs by $\bar Q^{(1)}$ cancel out in the supersymmetric index.

We consider one final example that will reappear in a three-dimensional problem in section~\ref{sec:XYZ}. We introduce three chiral multiplets $\phi_1,\phi_2,\phi_3$ coupled to three Fermi multiplets $\eta_1,\eta_2,\eta_3$ with $J$-term superpotentials
\be
J_1 = \phi_2\phi_3 \qquad J_2 = \phi_3\phi_1 \qquad J_3 = \phi_1\phi_2 \,.
\ee
This model in fact arises from a supersymmetric quantum mechanics with four supercharges with chiral multiplets $\Phi_1$, $\Phi_2$, $\Phi_3$ and cubic superpotential $W = \Phi_1\Phi_2\Phi_3$. Here, we regard it as an $\cN=(0,2)$ supersymmetric quantum mechanics with flavour symmetry
\begin{center}
\begin{tabular}{c | c c }
& $U(1)_T$ & $U(1)_A$  \\
\hline
$\phi_1$ & $1$ & $-1$  \\
$\phi_2$ & $-1$ & $-1$ \\
$\phi_3$ & $0$ & $2$
\end{tabular} .
\end{center}
Our notation and choice of charges is made with future applications in mind. It is straightforward to check that the supersymmetric index is $1$.

Introducing real mass parameters $m_T$ and $m_A$, there are six chambers $\mathfrak{c}_\al \subset \mathfrak{t}_f = \R^2$ labelled by sign vectors $\al = (\al_1,\al_2,\al_3)$. The sign vectors $(+++)$ and $(---)$ are not allowed as the mass parameters of all three chirals must sum to zero. Let us concentrate here on the chamber $(+-+)$ corresponding to mass parameters $0<m_A<m_T$. Following previous examples, we compute the cohomology of 
\be
\bar Q^{(1)} = \phi_1\phi_2 \eta_3 + \phi_2\phi_3 \eta_1 + \phi_3\phi_1 \eta_2
\ee
on states
\be
[\, \phi_1^{n_1} \, \bar\phi_2^{n_2} \, \phi_3^{n_3} \, ] \bar\eta_1^{s_1}\bar\eta_2^{s_2}\bar\eta_3^{s_3} \, .
\ee
where $n_1,n_2,n_3\geq0$ and $s_1,s_2,s_3 = 0,1$. We have chosen the Fock vacuum annihilated by $\psi_1, \psi_2,\psi_3$ and $\eta_1,\eta_2,\eta_3$ and assigned it Fermion number one.

This example is simple enough to compute representatives of cohomology classes directly. A more systematic method is to split the supercharge into three terms and apply the method spectral sequences to compute the cohomology of the total complex in steps. We summarize representatives of the remaining cohomology classes and their contribution to the supersymmetric index below,
$$
\begin{array}{l|l}
\quad \cH_{+-+} & \quad \cI_{+-+} \\ \hline
\left[\,\bar\phi_2^{n}\,\right] \quad & \quad +(qy)^{n+1} \\
\left[\,\phi_1\,\bar\phi_2^{n}\,\right]\bar\eta_{1} \quad & \quad -(qy)^{n+1}  \\
\left[\,\phi_3^{n}\,\right]\bar\eta_{1} \quad & \quad -(y^2)^{n+1}  \\
\left[\,\phi_3^{n+1}\,\right]\bar\eta_{1}\bar\eta_3 \quad & \quad +(y^2)^{n+1}  \\
\left[\,\phi_1^{n}\,\right]\bar\eta_{3} \quad & \quad -(qy^{-1})^{n+1}  \\
\left[\,\phi_1^{n+1}\,\right]\bar\eta_1\bar\eta_{3} \quad & \quad +(qy^{-1})^{n+1} \\
\left[\,1\,\right]\bar\eta_1\bar\eta_{3} \quad & \quad +1  
\end{array} \, ,
$$
where $n \geq 0$. All of the contributions to the supersymmetric index cancel in pairs except for the final line, reproducing the expected result $\cI = 1$. A similar analysis can be performed in the remaining chambers.

A general model consists of $N$ chiral multiplets $(\phi_j,\psi_j)$ and $k$ Fermi multiplets $(\eta^a,F^a)$ coupled by holomorphic superpotentials $J_a(\phi)$ and $E^a(\phi)$. The Hilbert space of supersymmetric ground states is the cohomology of 
\be
Q^{(1)} = \eta^a J_a(\phi) + \bar\eta_a E^a(\phi)
\ee
acting on the tensor product of supersymmetric ground states for the individual chiral and Fermi multiplets. 
This can be understood as a supersymmetric sigma model to $M = \mathbb{C}^N$ together with the $\mathbb{Z}_2$-graded hermitian vector bundle 
\be
F = K_{\C^N}^{1/2} \otimes  \frac{\wedge^\bullet f^* }{\sqrt{\det f^*}} \, ,
\ee
where $f$ denotes the odd rank-$k$ hermitian vector bundle on $\C^N$ with fibers spanned by the complex fermions $\eta^a$. The holomorphic differential $\delta = \bar Q^{(01)}$ is given by the sum of contraction with the holomorphic section $\eta^aJ_a(\phi)$ of $f$ and the wedge product with the holomorphic section $\bar\eta_aE^a(\phi)$ of $f^*$.


\subsubsection{Gauge Theory}
\label{sec:qm-gauge}

A vectormultiplet in $\cN=(0,2)$ supersymmetric quantum mechanics contains a gauge field $A_\tau$, a real scalar $\sigma$, and a real auxiliary field $D$, in addition to the complex fermions $\lambda$, $\tilde\lambda$. The real mass parameters introduced above can be regarded as coupling to a background vectormultiplet for the flavour symmetry $G_f$ and turning on a vacuum expectation value $m_f = \la \sigma_f\ra$ for the scalar component.

We now consider dynamical vectormultiplets for a gauge symmetry $G$. We focus on $G=U(1)$ and introduce $N$ chiral multiplets $(\phi_j,\psi_j)$ transforming with charge $Q_j$.  We also introduce a real FI parameter $\zeta>0$ and a supersymmetric Wilson line of charge $q$. These contribute $ \zeta D$ and $q(\sigma+iA_\tau)$ respectively to the lagrangian. Global anomaly cancelation requires
\be
q -\frac{1}{2}\sum_{j=1}^N Q_j \in \mathbb{Z} \, .
\label{eq:qm-anom}
\ee
This model will arise in computations of the supersymmetric twisted Hilbert space of three-dimensional gauge theories with $G=U(1)$ on $C = \mathbb{CP}^1$ in sections~\ref{sec:U(1)1/2}-\ref{sec:sqed}

To compute the supersymmetric ground states, we introduce a supersymmetric sigma model onto configurations minimizing the euclidean action in the `geometric regime'. We first note that the auxiliary field can be eliminated by its equation of motion to give $D = e^2(\mu - \zeta)$,
where $\mu = \sum_jQ_j|\phi_j|^2$ is the moment map for the $U(1)$ action on $\mathbb{C}^N$. The classical potential is then
\be
U(\sigma,\phi) = \sigma^2 \sum_{j=1}^N | Q_j \phi_j|^2 + \frac{e^2}{2}(\mu-\zeta)^2
\ee
Assuming $\zeta >0$, the classical potential is minimized by configurations 
\be
\mu-\zeta = 0 \qquad \sigma = 0
\ee
modulo constant $U(1)$ gauge transformations. At frequencies much smaller than $e^2\sqrt{\zeta}$, the system can be described by a supersymmetric sigma model to the K\"ahler quotient 
\be
M = \mu^{-1}(\zeta) / U(1)  \, .
\ee
Since the dependence on $e^2$ and $\zeta$ is exact, provided the quotient $M$ is smooth we expect the supersymmetric sigma model to exactly capture the space of supersymmetric ground states in the supersymmetric gauge theory. Note that for the quotient $M$ to be smooth, we require
\be
Q_j = \begin{cases}
+1 & \text{for} \quad j=1,\ldots,k  \\
-1 & \text{for} \quad j=k+1,\ldots,N \, ,
\end{cases}
\ee
such that $M$ is the total space of the line bundle $\cO(-1)^{N-k} \to \mathbb{CP}^{k-1}$. 

Let us specialize here to the compact case with $N$ chiral multiplets of charge $Q_j = +1$. The chiral multiplets transform in the fundamental representation of the $G_f = PSU(N)$ flavour symmetry. In this case, we find a supersymmetric sigma model to $M = \mathbb{CP}^{N-1}$. In addition, there is is a hermitian line bundle $F$ with the following contributions:
\begin{itemize}
\item A contribution $K_{\mathbb{CP}^{N-1}}^{1/2} = \cO(-\frac{N}{2})$ from quantizing chiral multiplet fermions.
\item A contribution $\cO(q)$ from the supersymmetric Wilson line.
\end{itemize}
The anomaly cancellation condition~\eqref{eq:qm-anom} ensures that the combination $F = \cO(q-\frac{N}{2})$ is well-defined. 

We can further introduce real mass parameters for the chiral multiplets $m_f = (m_1,\ldots,m_N)$ with $\sum_jm_j = 0$, which parametrize $\mathfrak{t}_f = \mathbb{R}^{N-1}$. In the supersymmetric sigma model, the real mass parameters introduce a real superpotential $h_f : \mathbb{CP}^{N-1} \to \R$ given by the moment map for the $U(1)_{m_f}$ isometry of $\mathbb{CP}^{N-1}$ generated by $m_f$. 

Since the target of the supersymmetric sigma model is compact, the spectrum is always gapped and the space of supersymmetric ground states is constant over the whole parameter space $\mathfrak{t}_f = \mathbb{R}^{N-1}$. In particular, at $m_f = 0$ the space of supersymmetric ground states can be identified with the Dolbeault cohomology
\be
\cH = H_{\bar\partial}^{0,\bullet}(M,F) \, .
\ee
We introducing complex parameters $(x_1,\ldots,x_N)$ with $\prod_j x_j =1$ parametrizing the complexified maximal torus of $G_f = PSU(N)$ to keep track of the  grading  by flavour symmetry. Then
\be
\cH = \begin{cases}
S^{q-\tfrac{N}{2}} \left( \bigoplus_{j=1}^N x^{-1}_j\C  \right) & \quad q \geq \tfrac{N}{2}  \\
\emptyset & -\tfrac{N}{2} < q< \tfrac{N}{2} \\
S^{-q-\tfrac{N}{2}} \left( \bigoplus_{j=1}^N x_j\C  \right)& \quad q \leq -\tfrac{N}{2}  \, ,
\end{cases}
\ee
corresponding to symmetric powers of the fundamental and anti-fundamental representations of $G_f = PSU(N)$. 

The supersymmetric index may be computed independently by localization in the supersymmetric gauge theory~\cite{Hori:2014tda}. This computation reproduces the characters of the above representations of $G_f = PSU(N)$,
\bea
I & = \oint_\Gamma \frac{dz}{z} z^q\prod_{j=1}^N \frac{(zx_j)^{1/2}}{1- z x_j} \\
  & = \begin{cases}
\chi_{S^{q-\frac{N}{2}}(\C^N)^*}( x_1,\ldots,x_N ) & \quad q \geq \tfrac{N}{2}  \\
0 & -\tfrac{N}{2} < q< \tfrac{N}{2} \\
\chi_{S^{-q-\frac{N}{2}}\C^N}( x_1,\ldots,x_N )& \quad q \leq -\tfrac{N}{2} \, .
\end{cases}
\eea
The contour $\Gamma$ surrounds the poles at $z = x_j^{-1}$ for all $j=1,\ldots,N$. The contour integral expression for the supersymmetric index coincides with the computation of the holomorphic Euler character $\chi(M,F)$ using the Hirzebruch-Riemann-Roch theorem.


\section{Three-Dimensional Setup}
\label{sec:3d-setup}

We now consider 3d $\cN=2$ theories with a topological twist on the product of a real line $\R$ and a compact connected Riemann surface $C$ of genus $g$. This setup falls into the general class of supersymmetric backgrounds introduced in~\cite{Closset:2017zgf}. The related supersymmetric background on $S^1 \times C$ has been further studied in the context of the supersymmetric twisted index in~\cite{Benini:2015noa,Benini:2016hjo,Closset:2016arn}.

This setup preserves the same supersymmetry as an $\cN=(0,2)$ supersymmetric quantum mechanics on $\mathbb{R}$ of the type considered in section~\ref{sec:susy-qm}. In this section, we will explain how the 3d supermultiplets decompose into those of the supersymmetric quantum mechanics and introduce an effective description that captures the twisted Hilbert space of supersymmetric ground states. The supersymmetric twisted Hilbert space is then computed explicitly in a number of examples in section~\ref{sec:examples}.


\subsection{Decomposing Supermultiplets}
\label{sec:decomp}

\subsubsection{Chiral Multiplets}
\label{sec:3d-chirals}

Let us first consider 3d chiral multiplets $\Phi$ of $R$-charge $r$ transforming in a unitary representation $R_f$ of a flavour symmetry $G_f$. 
We can introduce two deformation parameters associated to the flavour symmetry that are compatible with the topological twist on $C$:
\begin{itemize}
\item A holomorphic $G_f$ bundle $E_f$ on $C$ in the representation $R_f$.
\item Real mass parameters $m_f \in \tf_f$.
\end{itemize}
As explained below, these deformation parameters can be understood as vacuum expectation values for a background vectormultiplet for $G_f$.

Performing the topological twist on $C$, the three-dimensional chiral multiplet decomposes into the following supermultiplets in supersymmetric quantum mechanics:
\begin{itemize}
\item A 1d $\cN=(0,2)$ chiral multiplet $(\phi,\psi)$ valued in smooth sections of $E_\Phi$,
\item A 1d $\cN=(0,2)$ Fermi multiplet $(\eta,F)$ valued in smooth sections of $\Omega^{0,1}_C \otimes E_\Phi$, 
\end{itemize}
where
\be
E_\Phi := K_C^{r/2} \otimes E_f
\ee
and $K_C$ is the canonical bundle on $C$. This may require a choice a spin structure on $C$, and different choices are related by tensoring $E_f$ with a flat line bundle on $C$. The holomorphic bundle $E_\Phi$ inherits a hermitian metric from that on the canonical bundle on $C$ and the hermitian metric on the vector space of the unitary representation $R_f$.

In addition to the standard kinetic term contributions in the supersymmetric quantum mechanics from these supermultiplets, there is an $E$-type superpotential for the Fermi multiplet, 
\be
\mathcal{E}= \bar D \phi \, ,
\label{eq:e-sp}
\ee
where $\bar D$ denotes the holomorphic structure on $E_\Phi$. Note that the $E$-term superpotential transforms in the same way as the Fermi multiplet $(\eta,F)$, as required for supersymmetry. The $E$-term contribution to the supercharge $\bar Q$ is proportional to
\be
\int_C  \bar\eta \wedge \mathcal{E}  = \int_C   \bar\eta \wedge \bar D \phi \, ,
\ee
where contraction using the hermitian metric on $E_\Phi$ is understood. The choice of holomorphic structure $\bar D$ on $E_\Phi$ is therefore a $B$-type parameter in the supersymmetric quantum mechanics.

The role of the $E$-type superpotential is to provide kinetic terms with derivatives along $C$ in the supersymmetric quantum mechanics. In particular, it contributes terms to the Lagrangian of the supersymmetric quantum mechanics proportional to
\bea
|E|^2 & =  \int_C  D \bar\phi  \wedge * \bar D \phi \\
\mathrm{Re} \; \bar \eta \frac{\partial E}{\partial \phi }\psi  & =  \mathrm{Re} \; \int_C \bar\eta \wedge \bar D \psi \, ,
\label{eq:e-pot}
\eea
where again contraction using the hermitian metric on $E_\Phi$ is understood. 

Finally, real mass parameters generate the real superpotential
\be
h_f = \int_C * \,  ( m_f \cdot \mu_f ) \, ,
\label{eq:3d-realmass-superpotential}
\ee
where $\mu_f \in \tf^*_f$ is the moment map for the action of $G_f$ on the unitary representation $R_f$. In the presence of real mass parameters, the supercharges of the supersymmetric quantum mechanics are formally conjugated by the exponential factor $e^{h_f}$, as in section~\ref{sec:geometric-model}. This is an $A$-type deformation of the supersymmetric quantum mechanics.

\subsubsection{3d Superpotentials}

The above model can be deformed by a 3d superpotential $W(\Phi)$ preserving the $R$-symmetry used to perform the topological twist on $C$. A superpotential will place restrictions on the flavour symmetry $G_f$ and therefore the allowed background vector bundle $E_f$ and real mass parameters $m_f$. 

The superpotential must have $R$-charge $+2$ and will therefore transform as a section of the canonical bundle on the curve $C$ in the twisted theory. It introduces a $J$-term superpotential to the supersymmetric quantum mechanics,
\be
\mathcal{J}_W =   \frac{\partial W}{\partial \Phi} \, ,
\label{eq:J_W}
\ee
which transforms as a smooth section of $\Omega_C^{1,0}\otimes \bar E_\Phi$. This generates an additional contribution to the supercharge $\bar Q$ proportional to
\be
\int_C  \mathcal{J}_W \wedge \eta = \int_C \frac{\partial W}{\partial \Phi} \wedge \eta \, ,
\ee
where contraction using the hermitian metric on $E_\Phi$ is understood. The complex parameters in the 3d superpotential therefore become $B$-type deformation parameters in the supersymmetric quantum mechanics.

\subsubsection{Gauge Theory}

Now consider a 3d $\cN=2$ vectormultiplet for a compact group $G$. The bosonic fields in the vectormultiplet consist of the connection $A_\mu$, a real scalar $\sigma$, and a real auxiliary field $D$. Performing the topological twist on $C$, the 3d $\cN=2$ vectormultiplet decomposes into the following multiplets in the supersymmetric quantum mechanics:
\begin{itemize}
\item A 1d $\cN=(0,2)$ vectormultiplet consisting of $A_\tau$, $\sigma$, and auxiliary field $D^{\mathrm{1d}} := D + * \, F$ where $*$ is the Hodge operator on $C$.
\item A 1d $\cN=(0,2)$ chiral multiplet valued with complex scalar component given by the $(0,1)$-form component of the connection $\bar D$  on $C$.
\end{itemize}
In the supersymmetric quantum mechanics, the vectormultiplet is associated to the infinite-dimensional group $\cG$ of gauge transformations $g : C \to G$. 

Let us consider a dynamical vectormultiplet with gauge symmetry $G$ together with chiral multiplets $\Phi$ of $R$-charge $r$ transforming in a unitary representation $R$ of $G$. We suppose the chiral multiplets transform in a unitary representation $R_f$ of a residual flavour symmetry $G_f$. 

We can again introduce deformation parameters $m_f$ and $E_f$ associated to the flavour symmetry $G_f$. In the supersymmetric quantum mechanics, the chiral multiplets decompose as above with
\be
E_\Phi = K_C^{r/2} \otimes E \otimes E_f
\ee
where $E$, $E_f$ are the holomorphic vector bundles associated to the representations $R$, $R_f$. The parameters $m_f$, $E_f$ can in fact be understood as vacuum expectation values for a background vectormultiplet for the flavour symmetry $G_f$:
\begin{itemize}
\item The mass $m_f = \la \sigma_f\ra$ is a vacuum expectation for the 1d vectormultiplet. 
\item The holomorphic vector bundle $E_f$ is a vacuum expectation value for the 1d chiral multiplet $\bar D_f$. 
\end{itemize}
Preserving both supercharges of the supersymmetric quantum mechanics requires $D_f^{\mathrm{1d}} = 0$ and therefore we should also turn on a compensating auxiliary field $D_f = - * F_f$ given by the curvature of $E_f$.

If $G$ contains abelian factors, there is a topological flavour symmetry and we can turn on a vacuum expectation value for a background vectormultiplet for this symmetry. For example, if $G=U(1)$ there is a topological flavour symmetry $U(1)_T$ and we can introduce a real FI parameter $m_T = \zeta$ and holomorphic line bundle $L_T$. It is useful to note that the contribution to the Lagrangian of the supersymmetric quantum mechanics from a real FI parameter is
\be
- \frac{\zeta}{2\pi} \int_C *\, D  = - \frac{\zeta}{2\pi} \int_C  *\, D^{1\mathrm{d}}  +  \zeta \fm \, ,
\ee
where 
\be
\fm = \frac{1}{2\pi}\int_C F \, .
\ee
This can be regarded as an FI term in the supersymmetric quantum mechanics with parameter $\zeta_{1\mathrm{d}} = \mathrm{Vol_C} \zeta$ plus an additional contribution proportional to the flux. 

Finally, for each abelian or simple factor in $G$, we can introduce a supersymmetric Chern-Simons term. For example, the contribution to the supersymmetric quantum mechanics from a $G=U(1)$ Chern-Simons term at level $k$ is 
\be
 \frac{k}{2\pi} \int_C (\sigma + i A_\tau) F - \frac{k}{2\pi} \int_C  * \sigma D^{1d} \, .
\label{eq:cs-lag}
\ee
In general, we can introduce various mixed Chern-Simons contributions between gauge, flavour and R-symmetries.

\subsection{Effective Quantum Mechanics}
\label{sec:eff-qm}

In the previous section, we rephrased the topological twist of a 3d $\cN=2$ theory on $\mathbb{R} \times C$ as an infinite dimensional supersymmetric quantum mechanics on $\mathbb{R}$. We now introduce an effective supersymmetric quantum mechanics that captures the space of supersymmetric ground states. We focus exclusively on regimes where there are only `Higgs branch' vacua in the sense of \cite{Intriligator:2013lca}. In this case, the effective supersymmetric quantum mechanics is a sigma model onto the moduli space of vortex like configurations that minimize the effective euclidean action, following the philosophy of~\cite{Bershadsky:1995vm,Kapustin:2006pk,Kapustin:2006hi}

\subsubsection{Chiral Multiplets}

Let us return to the model with chiral multiplets transforming a unitary representation $R_f$ of a flavour symmetry $G_f$ from section~\ref{sec:3d-chirals}. The euclidean action is minimized by time-independent configurations that minimize of the potential of the supersymmetric quantum mechanics
\be
U  =  \int_C  \| \bar D \phi \|^2 : = \int_C D \bar\phi   \wedge * \bar D \phi \, ,
\ee
which is induced by the $E$-type superpotential term in equation~\eqref{eq:e-pot}. Such configurations therefore satisfy $\bar D \phi = 0$. In addition, time-independent solutions of the Fermi multiplet equations of motion obey $\bar D \bar \eta = 0$. 

We therefore consider an effective supersymmetric quantum mechanics consisting of a finite number of fluctuations:
\begin{itemize}
\item Chiral multiplets $(\phi,\psi)$ valued in $H^0(E_\Phi)$
\item Fermi multiplets $(\eta, F)$ valued in $H^1(E_\Phi)$ .
\end{itemize}
Let us define the number of chiral and Fermi multiplet fluctuations by
\be
n_C := h^0(E_\Phi) \qquad n_F :=  h^1(E_\Phi) \, .
\ee
The difference is determined by the Riemann-Roch theorem,
\be
n_C - n_F = c_1(E_\Phi) - \mathrm{rk}(E_\Phi) (1-g)  \, ,
\ee
and depends on the background holomorphic vector bundle only through $c_1(E_\Phi)$ and $\mathrm{rk}(E_\Phi)$. In contrast, the individual numbers of fluctuations may depend on the particular choice of holomorphic vector bundle with these parameters.

Introducing mass parameters $m_f\in \tf_f$, we can quantize the chiral and Fermi multiplet fluctuations as in section~\ref{sec:qm-examples}. We obtain a supersymmetric twisted Hilbert space $\cH_\al$ in each chamber $\mathfrak{c}_\al \subset \mathfrak{t}_f$ where all of the fluctuations are massive. Furthermore, each $\cH_\al$ will jump as the holomorphic vector bundle $E_f$ is varied whenever the numbers $n_C$ and $n_F$ change.

A supersymmetric Chern-Simons term for the flavour symmetry descends to a Chern-Simons term in the supersymmetric quantum mechanics and shifts the flavour grading of supersymmetric ground states. For example, for $G_f = U(1)$ the contribution~\eqref{eq:cs-lag} to the Lagrangian of the supersymmetric quantum mechanics is
\be
( m_f + i A_{f,\tau}) k \mathfrak{m}_f 
\ee
where $\mathfrak{m}_f = c_1(E_f)$  is the flavour flux on $C$. This shifts the flavour conserved charge by $J_f \to J_f + k\fm_f$. 
This can be further supplemented by a superpotential $W(\Phi)$. In this case, we will assume that we can first quantize the fluctuations as above and then implement the $J$-term superpotential~\eqref{eq:J_W} arising from $W(\Phi)$ in the supersymmetric quantum mechanics of these fluctuations. 

We construct explicitly the supersymmetric twisted Hilbert spaces $\cH_\al$ for a single chiral multiplet with a supersymmetric Chern-Simons term for the $G_f=U(1)$ flavour symmetry in section~\ref{sec:1chiral}.
We then proceed examples involving superpotentials in sections~\ref{sec:XY} and~\ref{sec:XYZ}.

\subsubsection{Abelian Gauge Theories}
\label{sec:abeliangauge}

We now return to supersymmetric gauge theory. At this point, we specialise to $G = U(1)$ with a supersymmetric Chern-Simons term at level $k$. We will introduce parameters $\zeta$, $L_T$ associated to the $U(1)_T$ topological symmetry and parameters $m_f$, $E_f$ associated to the  flavour symmetry $G_f$ acting on chiral multiplets.

As above, we now consider configurations minimizing the euclidean effective action. We first set $m_f = 0$ and then later turn the mass parameters back on in the effective supersymmetric sigma model description. Eliminating the auxiliary field in the vectormultiplet, the effective potential of the supersymmetric quantum mechanics is then given by
\bea
U_{eff} & =   \frac{e_{eff}^2}{2} \int_C \Big\| \frac{1}{e_{eff}^2}*F + \mu-\frac{1}{2\pi}\xi_{eff} \Big\|^2  \\
& +\frac{1}{2e_{eff}^2} \int_C \| \partial \sigma \|^2 + \int_C \| \bar D \phi \|^2 + \int_C  \| \sigma \cdot \phi \|^2 \, , 
\eea
where
\be
\xi_{eff} := \zeta_{eff} + \sigma k_{eff}
\label{eq:combo}
\ee
is a combination of the 1-loop quantum corrected parameters $\zeta_{eff}$, $k_{eff}$, which are piecewise constant functions of $\sigma$. In the absence of mass parameters $m_f$, we have $\zeta_{eff} = \zeta$. Finally, $\mu$ is the moment map for the action of $G = U(1)$ on the unitary representation $R$.

Configurations minimising the euclidean effective action are therefore solutions to
\begin{gather}
\label{eq:gen-vort}
 \frac{1}{e_{eff}^2} *F  + \mu  = \frac{1}{2\pi} \zeta_{eff}(\sigma) \\ 
 \partial \sigma = 0    \qquad  \bar D \phi = 0 \qquad \sigma \cdot \phi = 0 \nonumber
\end{gather}
modulo gauge transformations $g:C \to G$. In particular, the vectormultiplet scalar $\sigma$ is real and therefore must be a constant on $C$.

By an analysis similar to reference~\cite{Intriligator:2013lca}, the equations~\eqref{eq:gen-vort} admit a variety of different solutions depending on the supersymmetric Chern-Simons level $k$, the matter content and the value of $\zeta$. In this paper, we focus exclusively on regimes with `Higgs branch' solutions, characterised by $\sigma = 0$, a non-vanishing expectation value for $\phi$ and the gauge symmetry completely broken. We can then focus attention on solutions of the generalized vortex equations
 \be
\frac{1}{e_{eff}^2} *F  + \mu  = \frac{\zeta}{2\pi}   \qquad \bar D \phi = 0  \, ,
\label{eq:gen-vort-2}
\ee
modulo gauge transformations. We note that these equations play an important role in the $A$-twist of 2d $\cN=(2,2)$ gauge theories~\cite{Witten:1993yc,Morrison:1994fr}. 

The moduli space of solutions to the generalized vortex equations~\eqref{eq:gen-vort-2} is a disjoint union of components
\be
\cM = \bigcup_\fm \cM_\fm 
\ee
labelled by the flux
\be
\fm = \frac{1}{2\pi} \int_C F  \in \mathbb{Z} 
\ee
and each component $\cM_\fm$ is finite-dimensional. 
Since the dependence on the gauge coupling is exact, we expect the spectrum of supersymmetric ground states of $U(1)_T$ topological charge $\fm$ to be captured by an effective supersymmetric quantum mechanics that is a sigma model with target space $\cM_\fm$. 

The twisted Hilbert space of supersymmetric ground states has the form
\be
\cH = \bigoplus_{\fm \in \mathbb{Z}}q^\fm \cH_\fm \, ,
\ee
where we have introduced a parameter $q \in \mathbb{C}^*$ to measure the charge under the topological symmetry $U(1)_T$. In passing to the twisted index on $S^1 \times C$ with a circle of radius $\beta$, this parameter is identified with the exponentiated FI parameter, $q = e^{-2\pi \beta \zeta}$. 

The flavour symmetry $G_f$ descends to an isometry of $\cM_\fm$ and becomes a flavour symmetry in the effective supersymmetric quantum mechanics and each $\cH_\fm$ transforms as a virtual representation of $G_f$. Introducing real mass parameters generates a superpotential in the effective supersymmetric sigma model equal to the moment map $h_f$ for the $U(1)_{m_f} \subset G_f$ isometry generated by $m_f$. This has the effect of formally conjugating the supercharges by the exponential of $h_f$. 

From the general structure of supersymmetric quantum mechanics summarized in section~\ref{sec:geometric-model}, we propose that the supersymmetric ground states with topological charge $\fm$ are computed by the $L^2$-cohomology
\be
\cH_\fm = H_{\bar\partial_{m_f}+\delta}^{0,\bullet} (\cM_\fm,\cF_\fm)
\label{eq:susy-H-general}
\ee
where $\cF_\fm$ is an appropriate $\mathbb{Z}_2$ graded hermitian vector bundle on $\cM_\fm$. The vector bundle $\cF_\fm$ receives contributions from the following sources:
\begin{itemize}
\item A universal contribution of the square root of the canonical bundle $\sqrt{K_{\cM_\fm}}$ from quantizing the fermions in chiral multiplets parametrizing $\cM_\fm$.
\item There are contributions from Fermi multiplet zero modes, which are solutions to $D \eta = 0$ in the background of a solution to the vortex equations. They transform as sections of a holomorphic vector bundle $\mathfrak{f}$ over the moduli space $\cM_\fm$. There is then a  contribution to $\cF_\fm$ from quantizing these fluctuations,
\be
\bigoplus_{i} (-1)^i \frac{\wedge^i \mathfrak{f^*}}{\sqrt{\mathrm{Det}(\mathfrak{f^*})}}  \, .
\ee

\item A supersymmetric Chern-Simons term at level $k$ for gauge symmetry $G=U(1)$ contributes an additional factor $K_{\cM_\fm}^{-k}$~\cite{Collie:2008mx}. 
\item A background holomorphic line bundle $L_T$ for the topological flavour symmetry on $C$ induces a line bundle $\widetilde L_T$ on the moduli space $\cM_\fm$. This corresponds to an electric impurity in the language of~\cite{Tong:2013iqa}. Alternatively, the holomorphic line bundle $\widetilde L_T$ can be constructed mathematically from the universal bundle $C\times \cM_\fm$ and Deligne pairing
\footnote{We provide some justification of these claims in the appendices. In appendix~\ref{app:vortex-symmetric}, we summarize the construction of the holomorphic line bundle $\widetilde L_T$ using the universal line bundle on $C \times \cM_\fm$ and Deligne pairing. In appendix~\ref{app:impurities}, we show that the curvature of this line bundle agrees with the `dirty connection' introduced in~\cite{Tong:2013iqa}.}.
\end{itemize}
Finally, as in the case of chiral multiplets, a 3d superpotential $W(\Phi)$ generates an additional contribution $\delta$ to the differential.

Let us briefly comment on the dependence on the mass parameters $m_f$, mirroring the discussion in supersymmetric quantum mechanics for section~\ref{sec:geometric-model}. If the moduli space $\cM_\fm$ is non-compact, the cohomology~\eqref{eq:susy-H-general} will yield a different result in each chamber $\mathfrak{c}_\al \subset \tf_f$ separated by walls where there are massless non-compact fluctuations. On the other hand, if $\cM_\fm$ is compact we may set $m_f = 0$ and identify the supersymmetric twisted Hilbert space with the hypercohomology
\be
\cH_\fm = H^{0,\bullet}_{\bar\partial +\delta }(\cM_\fm , \cF_\fm)  \, .
\ee
Finally, the background holomorphic vector bundle $E_f$ is a $B$-type deformation parameter in the effective supersymmetric quantum mechanics and therefore $\cM_\fm$, $\cF_\fm$ and $\cH_\fm$ may jump as this is deformed.

\subsection{Vortex Moduli Spaces}
\label{sec:cortex-moduli-space}

The structure of the moduli spaces $\cM_\fm$ depends intricately on the gauge and $R$-charges of the chiral multiplets, the genus $g$ of $C$, and the background holomorphic vector bundle $E_f$. In this section, we nevertheless attempt to make some general comments on their structure that will be used in examples in section~\ref{sec:examples}.

First, the vortex moduli space can be understood as an infinite-dimensional K\"ahler quotient. The group of gauge transformations $g:C \to G$ acts on the infinite-dimensional flat K\"ahler manifold $\mathbb{M}$ parametrized by pairs $(\bar D,\phi)$ with moment map 
\be
\mu_\cG := \frac{1}{e_{eff}^2}*F + \mu
\label{eq:inf-moment-map}
\ee
Imposing the complex equation $\bar D \phi = 0$ defines a K\"ahler submanifold $\mathbb{N} \subset \mathbb{M}$ with moment map given by the restriction of the above. The moduli space of solutions to the generalized vortex equations is then the quotient $\cM = \mu_\cG^{-1}(\zeta) / \cG$.

Following reference~\cite{Morrison:1994fr}, it will be useful to introduce an algebraic description of the moduli spaces $\cM_\fm$. This is obtained by imposing the complex equation $\bar D \phi = 0$ as above but replacing the real moment map equation by a stability condition on the pair $(\bar D,\phi)$ and dividing by complexified gauge transformations $g: C \to G_\C$.
A point in $\cM_\fm$ is then specified by the following data:
\begin{itemize}
\item A holomorphic line bundle $E$ of degree $\fm$.
\item A holomorphic section $\phi $ of $E_{\Phi}$ subject to a stability condition.
\end{itemize}
We will specify the relevant stability condition in examples in section~\ref{sec:examples}.

Finally, the algebraic description makes it clear that there is a holomorphic map
\be
j : \cM_\fm \longrightarrow J_C
\label{eq:abel-jacobi}
\ee
to the Picard variety parametrizing the holomorphic line bundle $E$, whose fibers are toric varieties. This holomorphic map is not generally surjective. However, when $\fm$ is sufficiently large it become a holomorphic fibration whose structure can be useful for computing the cohomology groups~\eqref{eq:susy-H-general}. We will study an explicit example of this phenomenon in section~\ref{sec:sqed}. If $g = 0$ then $J_C$ is a point and we recover the description of the vortex moduli space as a toric variety~\cite{Morrison:1994fr}.


\section{Examples}
\label{sec:examples}


\subsection{1 Chiral Multiplet}
\label{sec:1chiral}

We first consider a single chiral multiplet $\Phi$ of integer $R$-charge $r$. We introduce a real mass parameter $m_f$ and a background holomorphic line bundle $L_f$ for the $U(1)_f$ flavour symmetry. For later convenience, we define
\be
L_\Phi := K_C^{r/2} \otimes L_f \, .
\ee
If necessary, we regard the choice of spin structure on $C$ as a base point on the Jacobian parametrizing the line bundle $L_\Phi$.

The effective supersymmetric quantum mechanics has the following supermultiplets of charge $+1$ under $U(1)_f$:
\begin{itemize}
\item Chiral multiplets $(\phi,\psi)$ valued in $H^0(L_\Phi)$.
\item Fermi multiplets $(\eta,F)$ valued in $H^1(L_\Phi)$.
\end{itemize}
Let us denote the number of chiral and Fermi multiplet fluctuations by $n_C = h^0(L_\Phi)$ and $n_F = h^1(L_\Phi)$ respectively. The difference is fixed by Riemann-Roch,
\bea
n_C - n_F 
& = \fm_\Phi- g + 1 \\
& = \fm_f + (r-1)(g-1) .
\eea
where $\fm_f$, $\fm_\Phi$ are the degrees of $L_f$, $L_\Phi$ respectively. 

For extreme values of the degree, the numbers of fluctuations are fixed
\bea
\fm_\Phi > 2g-2 \quad & \Rightarrow \quad 
\begin{cases}  
n_C = \fm_\Phi - g +1 \\
n_F = 0
\end{cases} \\
\fm_\Phi <0 \quad & \Rightarrow \quad 
\begin{cases} 
n_C = 0 \\
n_F = -\fm_\Phi+g-1 \, .
\end{cases}
\eea
However, in the intermediate region $0 \leq \fm_\Phi \leq 2g-2$, the individual numbers of fluctuations depend on the choice of background line bundle $L_\Phi$ and will jump at loci in the Jacobian parametrizing $L_\Phi$.

In quantizing these fluctuations, we will choose the Fock vacuum of Fermion number zero to be annihilated by the fermions $\psi$ and $\bar\eta$. This Fock vacuum then has charge $+\frac{1}{2}(n_C-n_F)$ under $U(1)_f$. There are two chambers to consider: $\mathfrak{c}_+ = \{m_f>0\}$ and $\mathfrak{c}_- = \{m_f<0\}$. Quantizing the fluctuations as in section~\ref{sec:qm-examples}, we find
\bea
\cH_+ &=  x^{\frac{n_C-n_F}{2}} \displaystyle\bigoplus_{j=0}^\infty x^j\left( \bigoplus_{p+q=j} S^p\mathbb{C}^{n_C} \otimes \wedge^q\C^{n_f}\right)  \\
\cH_-  & =  (-1)^{n_C-n_F} x^{-\frac{n_C-n_F}{2}} \displaystyle\bigoplus_{j=0}^\infty x^{-j}\left( \bigoplus_{p+q=j} S^p\mathbb{C}^{n_C} \otimes \wedge^q\C^{n_f}\right)  \, .
\label{eq:ChiralHilbertSpace}
\eea
This reproduces the expansions of the supersymmetric twisted index
\be
\cI = \left( \frac{x^{1/2}}{1-x} \right)^{n_C-n_F}
\label{eq:ChiralIndex}
\ee
in the regions $|x|<1$ and $|x|>1$ respectively. Note that the supersymmetric twisted index depends only on the difference $n_C-n_F$ and therefore the degree $\mathfrak{m}_\Phi$. On the other hand, $\cH_{\pm}$ will jump as the background line bundle $L_\Phi$ is varied.

Let us now consider the special case $C = \mathbb{CP}^1$. The background holomorphic line bundle is now fixed $L_\Phi = \cO(\fm_\Phi)$ with $\fm_\Phi = \fm_f - r$ and there are either chiral or Fermi multiplet fluctuations,
\bea
\fm_\Phi \geq 0\quad & \Rightarrow \quad 
\begin{cases} 
n_{C} = \fm_\Phi  +1 \\
n_F = 0
\end{cases} \\
\fm_\Phi <0 \quad & \Rightarrow \quad 
\begin{cases} 
n_C = 0 \\
n_F = |\fm_\Phi|-1 \, .
\end{cases}
\eea
In this case, the effective supersymmetric quantum mechanics has an additional flavour symmetry $U(1)_\ep$ transforming the homogeneous coordinates of $\mathbb{CP}^1$ by 
\be
(z,w) \to (\xi^{1/2}z,\xi^{-1/2}w) \, .
\ee
This induces an action on the chiral and Fermi multiplet fluctuations. For example, for $\fm_\Phi \geq 0$ the holomorphic sections $z^{\fm_\Phi-j} w^{j}$ transform with weight $\xi^{\frac{\fm_\Phi}{2}-j}$ for $j=0,\ldots,\fm_\Phi$. The vector space of chiral multiplet fluctuations therefore decomposes in the following way as a representation of $U(1)_\ep$,
\be
\xi^{\rho} \mathbb{C}^{\fm_\Phi+1} = \xi^{\frac{\fm_\Phi}{2}} \mathbb{C} \oplus \xi^{\frac{\fm_\Phi}{2}-1} \mathbb{C} \oplus \cdots \oplus \xi^{-\frac{\fm_\Phi}{2}} \mathbb{C} \, ,
\ee
where 
\be
\rho = \left( \frac{\fm_\Phi}{2},\frac{\fm_\Phi-1}{2},\ldots,-\frac{\fm_\Phi}{2} \right)
\ee
is the appropriate Weyl vector. Similar arguments apply for the Fermi multiplet fluctuations when $\fm_\Phi < 0$.

In the region $m_f>0$, we find
\be
\cH_+ = \begin{cases}
x^{\frac{\fm_\Phi+1}{2}} \displaystyle\bigoplus\limits_{j=1}^\infty x^j \, S^j( \xi^\rho \C^{\fm_\Phi+1} ) &  \quad \fm_\Phi \geq0\\
x^{\frac{\fm_\Phi+1}{2}} \displaystyle\bigoplus\limits_{j=1}^{|\fm_\Phi|-1} x^j \wedge^j (\xi^\rho \C^{|\fm_\Phi|-1} ) & \quad \fm_\Phi<0\, .
\end{cases} 
\ee
The supersymmetric index is consequently
\be
\cI = 
\begin{cases}
\prod\limits_{j=0}^{\fm_\Phi}  \frac{(\zeta^{\frac{\fm_\Phi}{2} - j}x)^{1/2}}{1-\xi^{\frac{\fm_\Phi}{2}-j}x} &  \quad \fm_\Phi\geq 0 \\
\prod\limits_{j=0}^{|\fm_\Phi|-2}  \frac{1-\zeta^{\frac{|\fm_\Phi|}{2}-1-j}x}{(\xi^{\frac{|\fm_\Phi|}{2}-1 - j}x)^{1/2}} & \quad \fm_\Phi <  0 \, ,
\end{cases}
\ee
which can be combined into the uniform expression
\be
\frac{x^{\frac{\fm_\Phi}{2}}}{(\xi^{-\frac{\fm_\Phi}{2}}x,\xi)_{\fm_\Phi+1}}\, ,
\ee
where $(a,q)_n$ is the finite Q-Pochhammer symbol. This is in agreement with the 1-loop determinant from supersymmetric localization~\cite{Benini:2015noa}.


\subsection{$XY$ Model}
\label{sec:XY}

We now consider a pair of chiral multiplets $\Phi_1$, $\Phi_2$ with a quadratic superpotential $W = \Phi_1\Phi_2$ and complementary $R$-charges $r_1 + r_2 = 2$. There is a $U(1)_f$ flavor symmetry under which $\Phi_1$ and $\Phi_2$ have charges $Q_1 =+1$ and $Q_2=-1$ respectively and we introduce a corresponding real mass parameter by $m_f$. We will focus here on $\mathfrak{c}_+ = \{m_f>0\}$ since the opposite chamber can be obtained by interchanging $\Phi_1$ and $\Phi_2$.

We introduce a background line bundle $L_f$ of degree $\fm_f$ for the $U(1)_f$ flavour symmetry and define $L_{\Phi_j} := K_C^{r_j/2} \otimes L^{Q_j}$. The chiral multiplet $\Phi_j$ contributes $n_{j,c} = h^0(L_{\Phi_j})$ chiral multiplets and $n_{j,f} = h^1(L_{\Phi_j})$ Fermi multiplets to the supersymmetric quantum mechanics. Combining Serre duality with $r_1 + r_2 = 2$, we have
\be
n_1 := n_{1,c} = n_{2,f} \qquad n_2 := n_{2,c} = n_{1,f} 
\ee
and furthermore from Riemann-Roch
\bea
n_{1} - n_{2} & = (r_1-1)(g-1) + \fm_f  \\
& = - (r_2-1)(g-1) + \fm_f \, .
\eea
Let us denote the chiral multiplet fluctuations by $\phi_{1,a}$, $\phi_{2,a'}$ and the Fermi multiplet fluctuations by $\eta_{1,a'}$, $\eta_{2,a}$ where $a = 1,\ldots, n_{1}$ and $a' = 1,\ldots, n_{2}$. The superpotential $W = \Phi_1\Phi_2$ induces the following $J$-type superpotentials 
\be
J_{\eta_{1,a'}} =  \phi_{2,a'} \qquad
J_{\eta_{2,a}} =  \phi_{1,a} \, .
\ee
for the Fermi multiplets in the supersymmetric quantum mechanics.

In quantizing these fluctuations, we choose the Fock vacuum with Fermion number zero to be annihilated by the fermions $\psi_{1,a}$, $\psi_{2,a'}$ and $\eta_{1,a'}$, $\eta_{2,a}$. This Fock vacuum is uncharged under $U(1)_f$. The supersymmetric twisted Hilbert space is then the cohomology of the supercharge
\bea
\bar Q^{(1)} & = \sum_{a' =1}^{n_2}   J_{\eta_{1,a'}} \eta_{1,a'} + \sum_{a =1}^{n_1} J_{\eta_{2,a}} \eta_{2,a}  \\
& = \sum_{a' =1}^{n_2}   \phi_{2,a'}\eta_{1,a'} + \sum_{a =1}^{n_1}  \phi_{1,a}\eta_{2,a} 
\eea
acting on wavefunctions
\be
\Big[ \, \prod_{a=1}^{n_1} \phi^{k_a}_{1,a} \prod_{a' = 1}^{n_2} \bar\phi^{k_{a'}}_{2,a'} \, \Big] \prod_{a'=1}^{n_2} \bar\eta^{s_{a'}}_{1,a'} \prod_{a=1}^{n_1} \bar\eta^{s_a}_{2,a}  \, ,
\ee
where $k_a,k_{a'} \in \mathbb{Z}_{\geq 0}$ and $s_1,s_2 \in \{0,1\}$ and we use the notation for representatives of supersymmetric ground states introduced in section~\ref{sec:qm-examples}. The result is essentially $n_1+n_2$ copies of the supersymmetric quantum mechanics considered there consisting of a chiral multiplet $\phi$ and a Fermi multiplet $\eta$, with superpotential $J=\phi$. Here, $n_1$ chiral multiplets have positive real mass and $n_2$ chiral multiplets have negative real mass.

There is a single supersymmetric ground state in the cohomology of $\bar Q^{(1)}$ with $k_a=k_{a'}=0$, $s_a=0$ and $s_{a'}=1$ for all $a=1,\ldots,n_1$ and $a'=1,\ldots,n_2$. This corresponds to the fermions $\prod_{a'}\bar\psi_{2,a'} \prod_{a} \bar\eta_{1,a'}$ acting on our choice of Fock vacuum. The supersymmetric ground state has vanishing $U(1)_f$ charge and Fermion number zero, and so we have $\cH_+ = \C$.

The supersymmetric index computed using localization is a product of contributions from the chiral multiplets $\Phi_1$ and $\Phi_2$,
\be
\cI = \left(\frac{x^{1/2}}{1-x}\right)^{(r_1-1)(g-1) + \fm}\left(\frac{x^{-1/2}}{1-x^{-1}}\right)^{(r_2-1)(g-1) - \fm} = (-1)^{(r_2-1)(g-1) - \fm} \, ,
\ee
which is independent of the three-dimensional superpotential $W = \Phi_1\Phi_2$. This agrees with the single supersymmetric ground state up to a sign related to our choice of Fermion number for the Fock vacuum.


\subsection{$XYZ$ Model}
\label{sec:XYZ}

Now consider three chiral multiplets $\Phi_1$,$\Phi_2$,$\Phi_3$ with cubic superpotential $W = \Phi_1\Phi_2\Phi_3$ and R-charges such that $\sum_{j=1}^3 r_j=2$. The flavour symmetry is
\be
G_f = ( \prod_{j=1}^3 U(1)_j )  / U(1)_D
\ee
where $U(1)_D$ is the diagonal combination. 

It is convenient to represent the charges of the chiral multiplets $\Phi_i$ under $U(1)_j$ by a flavour charge matrix $Q^i{}_j$ satisfying the following constraints:
\begin{itemize}
\item $\sum_{i=1}^3 Q^i{}_j = 0$ ensures that the superpotential has zero charge under $U(1)_j$;
\item $\sum_{j=1}^3 Q^i{}_j = 0$ ensures no fields are charged under the diagonal combination $U(1)_D$; 
\item $\mathrm{rk}(Q) = 2$ ensures the full flavour symmetry is manifest.
\end{itemize}
A simple choice of charge matrix is
\be
Q = \begin{pmatrix}
$1$ & $-1$ & 0 \\
$0$ & $1$ & $-1$ \\
$-1$ & $0$ & $1$
\end{pmatrix} \, .
\ee
We can turn on corresponding real mass parameters $m_j$ such that $\sum_{j=1}^3m_j=0$.

Furthermore, we can introduce background line bundles $L_j$ for each $U(1)_j$ such that
\be
L_1 \otimes L_2 \otimes L_3 = \cO_C\, .
\label{eq:XYZ-bundles}
\ee
and define $L_{\Phi_i} = K_C^{r_j/2} \otimes \prod_j L_j^{Q^i{}_j}$.
The chiral multiplet $\Phi_i$ then leads to $n_{i,c} = h^0(L_{\Phi_i})$ chiral multiplet fluctuations and $n_{i,f} = h^1(L_{\Phi_i})$ Fermi multiplet fluctuations. From Riemann Roch, these numbers satisfy
\be
n_{i,c}-n_{i,f} = (r_i-1)(g-1) + \sum_j Q^i{}_j \fm_j
\ee
and therefore
\be
\sum_i (n_{i,c}-n_{i,f}) = 1-g\, ,
\ee
where $\fm_j$ denotes the degree of $L_j$. 

Combining Serre duality and equation~\eqref{eq:XYZ-bundles} we find
\be
H^0(C,L_{\Phi_1} \otimes L_{\Phi_2}) \cong H^1(C,L_{\Phi_3})^* \, .
\ee
In particular, we have $n_{1,c} + n_{2,c} = n_{3,f} $ together with cyclic permutations of this relation.
The superpotential then induces $J$-type superpotentials in the effective supersymmetric quantum mechanics associated to the natural maps
\be
H^0(C,L_{\Phi_1}) \otimes H^0(C,L_{\Phi_2}) \to  H^0(C,L_{\Phi_1} \otimes L_{\Phi_2}) \cong H^1(C,L_{\Phi_3})^*
\ee
and their cyclic permutations, which are given by multiplication of holomorphic sections. We remark that these multiplication maps have a rich structure and play a pivotal r\^ole in the theory of algebraic curves~\cite{arbarello}.

The structure of the effective supersymmetric quantum mechanics depends intricately on the flux $\fm$ and background line bundles $L_A$ and $L_T$. We will first consider the case $g > 0$ choosing for simplicity to turn off background line bundles for flavour symmetries, before returning to perform a more systematic analysis for $g=0$ in the presence of non-vanishing background fluxes.

\subsubsection{Example: $g\geq 1$}

For concreteness, let us specialize to $r_3 = 2$ and $r_1 = r_2 = 0$ and rename $\Phi_1 = X$, $\Phi_2 = Y$ and $\Phi_3 = Z$. Furthermore, let us remove the redundant $U(1)$ flavour symmetry and write the flavour symmetry as $G_f = U(1)_T \times U(1)_A$ with charges
\begin{center}
\begin{tabular}{c | c c }
& $U(1)_T$ & $U(1)_A$  \\
\hline
$X$ & $1$ & $-1$  \\
$Y$ & $-1$ & $-1$ \\
$Z$ & $0$ & $2$
\end{tabular} .
\end{center}
The assignment of charges has been selected for later comparison with $U(1)$ supersymmetric QED. 

Turning off background line bundles for flavour symmetries we have $L_X = L_Y = \cO_C$ and $L_Z = K_C$ and therefore we find a supersymmetric quantum mechanics with 
\bea
n_{c,X}=n_{c,Y} = n_{f,Z} = 1 \\
n_{f,X}=n_{f,Y} = n_{c,Z} = g 
\eea
and superpotentials
\be
J_{\eta_{x,\al}} = yz_\al \qquad J_{\eta_{y,\al}} = xz_\al \qquad J_{\eta_z} = x y
\ee
where $\al=1,\ldots,g$. As an aside, we mention that non-trivial background line bundles would necessarily introduce $J$-terms that are sums of monomials, complicating the computation of supersymmetric ground states. The special case $g=1$ has already appeared as an example of supersymmetric quantum mechanics in section~\ref{sec:qm-examples}.

With applications to 3d mirror symmetry in mind, we will consider the chamber $\mathfrak{c}_{+-+} = \{ 0 < m_A < m_T \}$ inside the space of mass parameters $\tf_f = \R^2$. The Hilbert space of supersymmetric ground states in this chamber should reproduce the expansion of supersymmetric twisted index
\be
\cI = y^{2g-2} \left[ \frac{(1-q y^{-1})(1-q^{-1} y^{-1})}{(1-y^2)} \right]^{g-1} 
\label{eq:index-XYZ}
\ee
for $|q|<|y|<1$, where we have introduced fugacities $q$ and $y$ for the flavour symmetries $U(1)_T$ and $U(1)_A$ respectively.

\begin{figure}[htp]
\centering
\includegraphics[height=2.25cm]{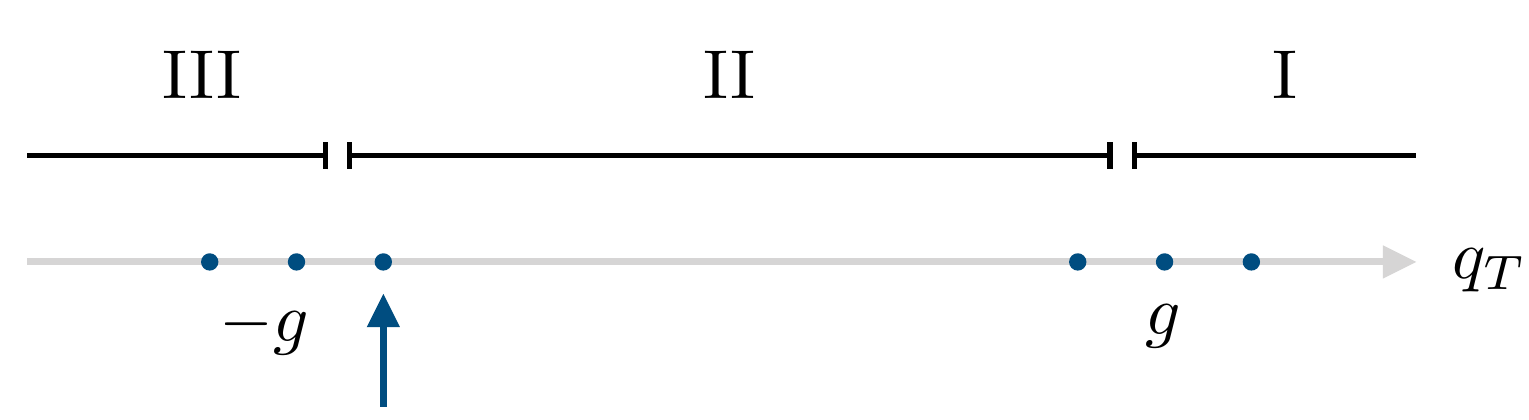}
\caption{We separate our discussion of the supersymmetric ground states for the XYZ model into three regions depending on the $U(1)_T$ flavour symmetry charge $q_T$. An arrow marks the point in region II where we obtain a complete result.}
\label{fig:xyz-regions}
\end{figure}

In order to enumerate the supersymmetric ground states in this model, it is convenient to choose the Fock vacuum annihilated by the fermions $\psi_{x}$, $\psi_y$, $\psi_{z,\al}$ and $\eta_{x,\al}$, $\eta_{y,\al}$, $\eta_z$ for all $\al = 1,\ldots,g$. This Fock vacuum is uncharged under the flavour symmetry and here we assign it Fermion number $(-1)^F = -1$ to match the supersymmetric twisted index. The supersymmetric ground states correspond to the cohomology of  
\be
\bar Q^{(1)} = \sum_{\al=1}^g yz_\al \eta_{x,\al} + \sum_{\al=1}^g xz_\al \eta_{y,\al} + x y \eta_{z}
\ee
acting on wavefunctions
\be
[ \, x^{n} \,  \bar y^{m} \, \prod_{\al=1}^g z_\al^{l_\al}  ] \prod_{\al=1}^g\bar\eta_{x,\al}^{s_\al} \prod_{\al=1}^g\bar\eta_{y,\al}^{t_\al} \bar\eta_z^{r} \, ,
\ee
where $n,m,l_\al \in \mathbb{Z}_{\geq 0}$ and $s_\al,t_\al,r \in \{0,1\}$. Since the flavour symmetry generators commute with the supercharge $\bar Q^{(1)}$, we can enumerate the supersymmetric ground states with fixed flavour charge separately. We find that the computation of the cohomology of $\bar Q^{(1)}$ splits into three qualitatively different regions depending on the $U(1)_T$ charge $q_T$. This is illustrated in figure~\ref{fig:xyz-regions}.

\vspace{10pt}

First, there are no supersymmetric ground states of charge $q_T \leq g$ and therefore the contributions to the supersymmetric twisted Hilbert space vanishes in region III,
\be
\cH_{(q_T)} = \emptyset \qquad q_T \leq g \, .
\ee
Next, we consider $q_T \geq g$ corresponding to region I and find the following representatives of the supersymmetric ground states with weights
\bea
\, & \big[ x^n \, \bar y^{q_T-1} \big]  \bar\eta_{x,i_1} \cdots \bar\eta_{x,i_n} \quad && : \quad  +(-1)^n q^{q_T} y^{q_T} \\
\, & \big[ x^{n+q_T-1}  \big]  \, \bar\eta_z \, \bar\eta_{x,i_1} ,\cdots,\bar\eta_{x,i_n}  \quad && : \quad  - (-1)^nq^{q_T} y^{-q_T}
\eea
for $\ell = 0 ,\ldots, n$ and therefore
\be
\cH_{(q_T)} = q^{q_T}( y^{q_T} - y^{-q_T} ) \otimes  \wedge^\bullet (\C^g) \qquad q_T \geq g\, .
\label{eq:Hilb.XYZ.g>0.1}
\ee
The twisted index should therefore vanish in region I due to the summation over contributions from an exterior algebra,
\be
\wedge^\bullet (\C^g) \longrightarrow \sum_{j=0}^g (-1)^j { g \choose j} = 0
\ee
for $g>0$. These results are consistent with the supersymmetric twisted index~\eqref{eq:index-XYZ}, which is a finite Laurent polynomial starting at order $\cO(q^{1-g})$ and ending at $\cO(q^{g-1})$.

In region II, where the supersymmetric twisted index is non-vanishing, the computation of the supersymmetric vacua is more intricate. For $g=1$, the computation is summarized as an example in section~\ref{sec:qm-examples}. The supersymmetric ground state representatives and their weights are
\bea
& \big[z^n\big] \bar\eta_{x} \qquad && - y^{2n+2}\\
& \big[z^n\big] \bar\eta_{x}\bar\eta_z \qquad && + y^{2n} 
\eea
for all $n\geq 0$ and thus
\be
\cH_{(0)}= (1-y^2) \bigoplus_{n\geq 0} y^{2n} \C  \qquad \text{if} \quad g = 1 \, .
\ee
This reproduces the supersymmetric index, which is $1$. 

For $g>1$, we have not obtained a systematic closed form for representatives of the supersymmetric ground states in region II. However, the sector $q_T = 1-g$ is a generalization of the above computation at $g=1$ and we find 
\bea
\big[ z_1^{n_1} &\cdots z_g^{n_g} \big] \bar\eta_{x_1} \cdots \bar\eta_{x_g}  \qquad && +(-1)^{g} q^{1-g}y^{g+1+2\sum_\al n_\al}   \\
\big[z_1^{n_1} &\cdots z_g^{n_g} \big] \bar\eta_{x_1} \cdots \bar\eta_{x_g}  \bar\eta_z  \qquad && -(-1)^{g} q^{1-g}y^{g-1+2\sum_\al n_\al}
\eea
for any $n_1,\ldots,n_g\geq 0$ and therefore
\be
\cH_{(1-g)} = (-1)^{g-1}q^{1-g} y^{g-1}(1-y^2) \, S^\bullet\left(y^2\C^g \right)  \, .
\label{eq:Hilb.XYZ.g>0.2}
\ee
This reproduces the correct contribution to the supersymmetric twisted index~\eqref{eq:index-XYZ} with $q_T = 1-g$,
\be
\cI_{(1-g)} = (-1)^{g-1}q^{1-g}\left(\frac{y}{1-y^2}\right)^{g-1} \, .
\ee

\subsubsection{Example: $g=0$}

We now consider $g=0$ and introduce non-trivial holomorphic line bundles $L_A = \cO(\fm_A)$ and $L_{T} = \cO(\fm_T)$ for the flavour symmetries such that
\bea
L_{\Phi_1} & = \mathcal{O}(\fm_T -\fm_A) \\
L_{\Phi_2} & = \mathcal{O}(-\fm_T -\fm_A) \\
L_{\Phi_3} & = \mathcal{O}(2 \fm_A-2) \, .
\eea
The supersymmetric twisted index is
\be
\mathcal{I} = \left( \frac{y}{1-y^2}\right)^{2 \fm_A-1} \left(\frac{q^{1/2}y^{-1/2}}{1-qy^{-1}} \right)^{\fm_T - \fm_A +1} \left( \frac{q^{-1/2}y^{-1/2}}{1-q^{-1}y^{-1}}\right)^{-\fm_T - \fm_A+1} \, .
\label{eq:XYZg0Index}
\ee

In contrast to $g>0$, for any given $(\fm_A, \fm_T)$ each 3d chiral multiplet leads to either chiral or Fermi multiplet fluctuations in the supersymmetric quantum mechanics, but not both. Now, a $J$-term contribution to the supercharge $\bar Q^{(1)}$ can only exist when two 3d chiral multiplets have chiral multiplet fluctuations and one has Fermi multiplet fluctuations. This happens when the fluxes that obey either 
\begin{enumerate}
\item $\fm_T \geq \fm_A \geq 1$ , or
\item $\fm_A \leq 0$ and $|\fm_T| \leq -\fm_A$ .
\end{enumerate}
Outside of these regions, the supersymmetric ground states are obtained simply by quantizing the chiral and Fermi multiplet fluctuations. We now spell out the supersymmetric ground states in two such regions needed to perform checks of mirror symmetry in section~\ref{sec:mirror}. As above, we consider the chamber $\mathfrak{c}_{+-+} = \{ 0 < m_A < m_T \}$ in the space of real mass parameters.

First, we consider fluxes obeying the constraints $\fm_A \geq 1$, $|\fm_T|< \fm_A$. Here, $L_{\Phi_1}$, $L_{\Phi_2}$ have negative degree and contribute only Fermi multiplet fluctuations, whereas $L_{\Phi_3}$ has non-negative degree and contributes only chiral multiplet fluctuations. There is no possibility of a $J$-term superpotential and therefore we find a tensor product of chiral and Fermi multiplet Fock spaces,
\be
\mathcal{H} = q^{\fm_T} y^{3\fm_A-2}S^{\bullet} (y^2 \mathbb{C}^{2 \fm_A-1}) \otimes \wedge^{\bullet} (q y^{-1}\mathbb{C}^{-\fm_T +\fm_A-1})  \otimes  \wedge^{\bullet} (q^{-1}y^{-1}\mathbb{C}^{\fm_T +\fm_A-1}) \, , 
\ee
reproducing the supersymmetric twisted index~\eqref{eq:XYZg0Index}. For future comparison with mirror symmetry, it is useful to expand this result in powers of $q$,
\bea
\mathcal{H} = &  y^{3\fm_A-2} S^{\bullet} (y^2 \mathbb{C}^{2 \fm_A-1}) \otimes \\ &\otimes \bigoplus^{-\fm_T+\fm_A-1}_{\fm=-\fm_T-\fm_A+1} q^{\fm_T+\fm} \left( \bigoplus_{i-k=\fm} \wedge^{i} (y^{-1} \mathbb{C}^{-\fm_T+\fm_A-1}) \otimes \wedge^{k} (y^{-1} \mathbb{C}^{\fm_T+\fm_A-1}) \right) \, .
\label{eq:XYZg0Hilb1}
\eea

Second, we consider $\fm_A \leq 0$, $\fm_T > - \fm_A$. Here, $L_{\Phi_1}$ has non-negative degree and contributes only chiral multiplet fluctuations, whereas $\Phi_2$ and $\Phi_3$ have negative degree and contribute only Fermi multiplet fluctuations. There are again no possible $J$-term superpotentials. We therefore have
\be
\mathcal{H} = q^{\fm_T} y^{3\fm_A-2} S^{\bullet} (qy^{-1}\mathbb{C}^{\fm_T -\fm_A+1})  \otimes  \wedge^{\bullet} (q^{-1}y^{-1}\mathbb{C}^{\fm_T +\fm_A-1}) \otimes \wedge^{\bullet} (y^2 \mathbb{C}^{-2 \fm_A+1})  \, ,
\ee
reproducing the supersymmetric twisted index~\eqref{eq:XYZg0Index}. Expanding again in $q$, this becomes
\bea
\mathcal{H} = &q^{\fm_T} y^{3\fm_A-2}  \wedge^{\bullet} (y^2 \mathbb{C}^{-2 \fm_A+1}) \otimes \\ &\otimes \bigoplus^{\infty}_{\fm=-\fm_T-\fm_A+1} q^{\fm} \left( \bigoplus_{k=0}^{\fm_T+\fm_A-1} S^{\fm-k} (y^{-1} \mathbb{C}^{\fm_T-\fm_A+1}) \otimes \wedge^{k} (y^{-1} \mathbb{C}^{\fm_T+\fm_A-1}) \right) \, .
\label{eq:XYZg0Hilb2}
\eea

\subsection{$U(1)_{\frac{1}{2}}+1$ Chiral Multiplet}
\label{sec:U(1)1/2}

We now consider supersymmetric Chern-Simons theory with $G = U(1)$ at level $k = +\frac{1}{2}$ together with a single chiral multiplet $\Phi$ of charge $+1$ and $R$-charge $+1$.

 In the supersymmetric quantum mechanics, the chiral multiplet $\Phi$ decomposes into a chiral multiplet and Fermi multiplet valued in sections of the holomorphic line bundle $L_\Phi$ and $\Omega_C^{0,1} \otimes L_\Phi$ respectively, where 
\be
L_\Phi = K_C^{1/2} \otimes L \, .
\ee
The supersymmetric quantum mechanics localizes to solutions of the following system of equations on $C$,
\be
\label{eq:vortex-U(1)-1/2}
\frac{1}{e_{eff}^2}*F + \bar\phi \cdot \phi  = \frac{1}{2\pi} \xi_{eff}(\sigma) \qquad  \bar D \phi = 0 \qquad \sigma \phi = 0  \, ,
\ee
where $\sigma$ is constant and
\be
\xi_{eff}(\sigma) = \begin{cases} \zeta + \sigma & \sigma > 0 \\ \zeta & \sigma < 0 \end{cases}
\ee
is the combination of effective parameters introduced in equation~\eqref{eq:combo}.

By integrating the first equation in~\eqref{eq:vortex-U(1)-1/2} over $C$, we find that there are `Higgs branch' vortex solutions with $\sigma = 0$ and $\phi$ non-vanishing provided the FI parameter obeys
\be
\zeta > \frac{4\pi^2 \fm}{e_{eff}^2 \mathrm{vol}_C}  \, ,
\label{eq:vortex-constraint}
\ee 
where
\be
\fm =  c_1(L) = \frac{1}{2 \pi} \int_{C} F 
\ee
and $\mathrm{Vol}_C$ is the volume of $C$. Here we are interested in the limit $e_{eff}^2 \to \infty$ where this condition is satisfied for any flux $\fm$ with $\zeta>0$. We can then introduce an effective description of the supersymmetric quantum mechanics as a sigma model onto the moduli space $\cM_\fm$ of solutions to the vortex equations
\be
\frac{1}{e_{eff}^2}*F + \bar\phi \cdot \phi  = \zeta \qquad  \bar D \phi = 0  \qquad \sigma = 0 \,
\label{eq:vortex-U(1)-1/2-reduced}
\ee
with flux $\fm$, modulo gauge transformations. 

As explained in section~\eqref{sec:eff-qm}, $\cM_\fm$ has a complex algebraic description in terms of the following holomorphic data:
\begin{itemize}
\item A holomorphic line bundle $L$ of degree $\fm$
\item A holomorphic section $\phi  \in H^0(L_\Phi)$ that does not vanish identically on $C$.
\end{itemize}
This is equivalent to a choice of effective divisor $D$ of degree
\bea
\fm_\Phi & := \deg(L_\Phi) \\
& = \fm + g - 1 \, ,
\eea
from which we can recover $L = \cO(D)$ and $\phi$ as the global holomorphic section with zeros on $D$. The moduli space of vortices can therefore be identified with the symmetric product parametrizing the effective divisor $D$,
\be
\cM_\fm = \begin{cases}
\mathrm{Sym}^{\fm_\Phi}C & \mathrm{if} \quad \fm_\Phi\geq0 \\
\emptyset & \mathrm{if} \quad \fm_\Phi < 0 \, .
\end{cases}
\ee
This is the statement that vortices in this theory have no internal moduli and symmetric product simply parametrizes the positions of the vortices on $C$.

For future reference, we note that there is a holomorphic map
\be
j : \cM_\fm \to J_C
\ee
to the Picard variety parametrizing the line bundle $L_\Phi$. The preimage of a point is the projective space of holomorphic sections, $j^{-1}(L_\Phi) = \mathbb{P}H^0(L_\Phi)$. The structure of this map is in general intricate since the dimension $h^0(L_\Phi)$ may jump at loci in the Jacobian. However, if $\fm_\Phi > 2g-2$ the dimension $h^0(L_\Phi) = \fm_\Phi - g +1$ is constant and this holomorphic map becomes a holomorphic fibration with fiber $\mathbb{CP}^{\fm_\Phi-g}$. Further details about vortices and symmetric products are given in appendix~\ref{app:vortex-symmetric}.

The contribution to the twisted Hilbert space of supersymmetric vacua from flux $\fm$ is captured by a supersymmetric sigma model to $\cM_\fm$ and therefore
\be
\cH_\fm =  \begin{cases}
H_{\bar\partial}^{0,\bullet} ( \cM_\fm , \cF_\fm) & \quad \fm \geq 1-g \\
\emptyset & \quad \fm < 1-g
\end{cases}
\label{eq:U(1)-hilb}
\ee
where $\cF_\fm$ is a hermitian line bundle on $\cM_{\fm}$~\footnote{An important consistency check is that the remaining massless fermions in chiral multiplets transform in the tangent bundle to $\cM_\fm$. In appendix~\ref{app:tangent}, we provide a short argument why this claim. We expect this to be important for future generalizations of this work.}. The line bundle $\cF_\fm$ receives contributions from the following sources:
\begin{itemize}
\item There is a universal contribution $K^{1/2}_{\cM_\fm}$ from quantizing fermions in chiral multiplets parametrizing $\cM_{\fm}$.
\item The supersymmetric Chern-Simons term at level $+\tfrac{1}{2}$ contributes a factor $\cK_{\cM_{\fm}}^{-1/2}$. 
\item Introducing a background line bundle $L_T$ for the topological symmetry induces a line bundle $\widetilde L_T$ on the moduli space $\cM_{\fm}$ in the supersymmetric quantum mechanics, as explained in section~\ref{sec:eff-qm}.
\end{itemize}
\noindent The universal contribution and that from the supersymmetric Chern-Simons term cancel out leaving $\cF_\fm = \widetilde L_T$.

For $\zeta < 0$ we find `topological' solutions to equations~\eqref{eq:vortex-U(1)-1/2} with $\phi = 0$, $\sigma = - \zeta$ and unbroken gauge symmetry as $e^2_{eff} \to 0$. We expect that the supersymmetric ground states are captured by an effective supersymmetric Chern-Simons theory in this regime and hope to return to this in future work~\footnote{We would like to thank Heeyeon Kim for discussions on this point}.

\subsubsection{Example: $g>0$}

The supersymmetric ground states in equation~\eqref{eq:U(1)-hilb} can be computed by applying the K\"unneth formula for the product of curves $C^{\fm_\Phi}$ and then imposing invariance under permutations to compute cohomology classes on the symmetric product regarded as a quotient $\sym^{\fm_\Phi}C = C^{\fm_\Phi} / S_{\fm_\Phi}$. 

This argument will rely on the following construction of the line bundle $\widetilde L_T$ associated to the $U(1)_T$ topological symmetry, which we expand upon in appendix~\ref{app:universal}. The first step is to construct a line bundle on the direct product $C^{\fm_\Phi}$,
\be
L_T^{\boxtimes \, \fm_\Phi} := \bigotimes_j \pi_j^*L_T \, ,
\label{eq:product-bundle}
\ee 
 where
\be
\pi_j : C^{\fm_\Phi} \to C
\ee
is the projection onto the $j$-th factor. This is invariant under permutations and descends to a line bundle $\widetilde L_T$ on the symmetric product $\sym^{\fm_\Phi}C$. In particular, this construction shows that $c_1(\widetilde L_T) = \fm_T \eta$ where $\fm_T$ is the degree of $L_T$ and $\eta \in H^{1,1}(\cM_\fm)$ is the class constructed from the K\"ahler form on $C$.

With the above construction in hand, we can proceed to compute the twisted Hilbert space following arguments in~\cite{Eriksson:2016wra}, but including higher degree cohomology. First, a short spectral sequence argument shows that the cohomology of the symmetric product in equation~\eqref{eq:U(1)-hilb} can be identified with the $S_{\fm_{\Phi}}$-invariant part of the cohomology $H^{0,\bullet}_{\bar\partial}(C^{\fm_\Phi},L_T^{\boxtimes \fm_\Phi})$. The latter can be computed using the K\"unneth decomposition, with the result
\be
H^{0,j}_{\bar\partial}(\sym^{\fm_\Phi}C,\widetilde L_T) = \mathrm{S}^{\fm_\Phi-j}H^{0}(L_T)  \otimes \wedge^j H^1(L_T) \, .
\ee
We therefore find
\be
\cH = \bigoplus_{\fm = 1-g}^\infty q^{\fm} \bigoplus_{i+j= \fm+g-1} S^i H^0(L_T) \otimes \wedge^jH^1(L_T) \, .
\label{eq:U(1)1/2HilbertSpace}
\ee
where the parameter $q$ is introduced to keep track of charge under the topological flavour symmetry and  we remind the reader that $\fm_\Phi = \fm + g -1$.

Introducing the notation $n_C := h^0(L_T)$ and $n_F := h^1(L_T)$ with
\be
n_C - n_F = \fm_T-g+1 \, ,
\ee
the supersymmetric twisted index can be computed from the graded trace over the twisted Hilbert space~\eqref{eq:U(1)1/2HilbertSpace} as follows,
\bea
\cI & = \sum_{\fm = 1-g}^\infty q^\fm { n_C-n_F +\fm - g \choose \fm-g+1} \\
 & = q^{1-g}  \left( \frac{1}{1-q} \right)^{\fm_T-g+1} 
\eea
This agrees with the contour integral formula from localization in the supersymmetric gauge theory~\cite{Benini:2016hjo,Closset:2016arn},
\be
\sum_{\fm \in \mathbb{Z}} (-q)^\fm \int_\Gamma \frac{dz}{2\pi iz} \frac{ z^{\fm+\fm_T}}{(1-z)^{\fm+g}} \, ,
\label{localization}
\ee
where the contour $\Gamma$ surrounds the pole at $z=1$, modulo an overall sign $(-1)^g$. The twisted supersymmetric index can be identified with the generating function for equivariant Euler characters
\be
I = \sum_{\fm = 1-g}^\infty(-q)^\fm\chi(\cM_\fm,\cF_\fm) \, .
\ee
From this perspective, the contour integral~\eqref{localization} from supersymmetric localization reproduces the computation of these equivariant Euler characters using Hirzebruch-Riemann-Roch~\cite{MACDONALD1962319}. In particular, the integration parameter can be identified with $z = e^{\eta}$ where $\eta \in H^{1,1}(\cM_\fm)$ is the class inherited from the K\"ahler form on $C$.

A simple example summarized in the introduction is $L_T = \cO_C$, where $n_C = 1$ and $n_F = g$ and therefore
\bea
\cH 
& = \bigoplus_{\fm = 1-g }^\infty q^\fm \bigoplus_{j=0}^{\fm+g-1} \wedge^{j}(\C^g)
\eea
with supersymmetric twisted index $I  = q^{1-g}(1-q)^{g-1}$.
Note that while the twisted Hilbert space contains an infinite number of supersymmetric ground states, the supersymmetric twisted index truncates to a finite number of terms for $g \geq 1$ due to the complete cancellation in the sum over the exterior algebra when $\fm >0$.

\subsubsection{Example: $g=0$}

Let us now analyse $g=0$ in more detail. The vortex moduli space $\cM_\fm = \mathbb{CP}^{\fm-1}$ is now the complex projective space parametrizing non-vanishing holomorphic sections of $L_\Phi = \cO(\fm-1)$ moduli  constant complex gauge transformations. This is consistent with the above construction since
\be
\mathrm{Sym}^{\fm-1} \mathbb{CP}^1 \cong \mathbb{CP}^{\fm-1} \,.
\ee
The supersymmetric quantum mechanics admits a finite-dimensional gauged linear sigma model description in terms of $\fm$ chiral multiplets of charge $+1$ under a $U(1)$ gauge symmetry. 

In this case the contributions to $\cF_\fm$ have a more straightforward interpretation:

\begin{itemize}
\item The contribution to the lagrangian of the supersymmetric quantum mechanics from a supersymmetric Chern-Simons term at level $k$ is
\be
\frac{k}{2\pi} (\sigma+iA_\tau) \int_C F =  (\sigma+iA_\tau) k \fm \, .
\ee
This is a supersymmetric Wilson line of charge $k \fm$ in the finite-dimensional gauged quantum mechanics for $\mathbb{CP}^{\fm-1}$ and so contributes the line bundle $\cO(k\fm)$. For a supersymmetric Chern-Simons term at level $k =+\tfrac{1}{2}$ we find a contribution $\cO(+\tfrac{\fm}{2})$.
\item Turning on $L_T = \cO(\fm_T)$ for the topological symmetry contributes
\be
\frac{1}{2\pi} (\sigma+ iA_\tau) \int_C F_T = (\sigma+ iA_\tau) \fm_T \, ,
\ee
which is a supersymmetric Wilson line of charge $\fm_T$ in the finite-dimensional gauged supersymmetric quantum mechanics and therefore the line bundle $\widetilde L_T = \cO(\fm_T)$ on $\mathbb{CP}^{m-1}$. 
\end{itemize}
\noindent Combining these contributions with the universal contribution $K^{1/2}_{\mathbb{CP}^{m-1}} = \cO(-\frac{m}{2})$ we find $\cF = \cO(\fm_T)$ in agreement with our previous arguments.

Including the parameter $\xi$ for the additional grading on $C = \mathbb{CP}^1$, the supersymmetric Hilbert space now has the form
\bea
\mathcal{H} & = \bigoplus_{\fm = 1}^\infty q^\fm H_{\bar{\partial}}^{0,\bullet} (\mathbb{CP}^{\fm-1}, \cO (\fm_T)) \\
& = \bigoplus\limits_{\fm = 1}^\infty q^\fm \begin{cases} 
 S^\tf(\xi^\rho \C^\fm) & \mathrm{if} \quad \fm_T  \geq0 \\
  \emptyset & \mathrm{if} \quad -\fm < \fm_T < 0 \\
S^{|\tf|-\fm}(\xi^\rho \C^\fm) & \mathrm{if} \quad \fm_T  \leq -\fm
\end{cases}
\, .
\eea

\subsection{$U(1)$ SQED}
\label{sec:sqed}

We now consider a $U(1)$ gauge theory with two chiral multiplets $\Phi$, $\tilde \Phi$ of charge $+1$, $-1$. The theory has both a $U(1)_T$ topological and $U(1)_A$ axial flavour symmetry and we can introduce corresponding background line bundles $L_T$ and $L_A$ of degrees $\fm_T$ and $\fm_A$. The charge assignments are summarized below,
\begin{center}
\begin{tabular}{c | c c c c}
& $U(1)_G$ & $U(1)_T$ & $U(1)_A$ & $U(1)_R$ \\
\hline
$\Phi$ & $1$ & $0$ & $1$ & $1$ \\
\hline
$\tilde \Phi$ & $-1$ & $0$ & $1$ & $1$ 
\end{tabular} .
\end{center}
The supersymmetric twisted index is given by
\bea
I & = \sum_{\fm\in\mathbb{Z}} (-\zeta)^\fm \int_\Gamma \frac{dx}{2\pi i\, x} x^{\fm_T} \left( \frac{x^\frac{1}{2} y^\frac{1}{2}}{1-xy} \right)^{\fm+\fm_A} \left( \frac{x^{-\frac{1}{2}} y^\frac{1}{2}}{1-x^{-1}y} \right)^{-\fm+\fm_A} \left(\frac{1-y^2}{(1-xy)(1-y/x)}\right)^g \\
& = (-1)^{\fm_T} \left( \frac{y}{1-y^2} \right)^{2\fm_A+g-1}  \left( \frac{\zeta^{1/2}y^{-1/2}}{1-\zeta y^{-1} } \right)^{\fm_T-\fm_A+1-g}  \left( \frac{\zeta^{-1/2}y^{-1/2}}{1-\zeta^{-1} y^{-1}} \right)^{-\fm_T-\fm_A+1-g} \, ,
\label{eq:sqed-index}
\eea
where the contour $\Gamma$ is specified by a Jeffrey-Kirwan prescription~\cite{Benini:2016hjo,Closset:2016arn}.

In the supersymmetric quantum mechanics the chiral multiplets $\Phi$, $\tilde \Phi$ decompose into chiral and Fermi multiplets transforming as scalar and $(0,1)$-form sections of the holomorphic line bundles
\be
L_\Phi := K^{1/2}_C\otimes L\otimes L_A \, ,\qquad L_{\tilde \Phi} := K^{1/2}_C \otimes L^{-1}\otimes L_A \, ,
\ee
of degrees
\be
\fm_\Phi = \fm + \fm_A  + g - 1 \, ,\qquad \fm_{\tilde \Phi} = - \fm + \fm_A  + g - 1 \, .
\ee
First setting the mass parameter for the axial flavour symmetry to zero, $m_A = 0$, the supersymmetric quantum mechanics localizes to solutions of the following system of equations on $C$,
\be
\frac{1}{e_{eff}^2}*F + \bar\phi \cdot \phi -\bar{\tilde\phi} \cdot \tilde\phi = \frac{\zeta}{2\pi} \qquad  \bar D \phi  = \bar D  \tilde\phi = 0 \qquad \sigma \phi  = \sigma \tilde\phi  = 0  \, ,
\label{eq:vortex-SQED}
\ee
where $\xi_{eff} = \zeta$ when the axial mass parameter vanishes. In the limit $e^2_{eff} \to \infty$ there are `Higgs branch' vortex solutions with $\sigma = 0$ whenever $\zeta \neq 0$. We will focus here on the regime $\zeta >0$.

We therefore consider an effective supersymmetric sigma model for each flux $\fm \in \mathbb{Z}$ whose target is the moduli space $\cM_\fm$ of solutions to equation~\eqref{eq:vortex-SQED} with $\zeta >0$. This moduli space has an algebraic description in terms of the following holomorphic data:
\begin{itemize}
\item A holomorphic line bundle $L$ of degree $\fm$.
\item A pair of holomorphic sections $\phi  \in H^0(C,L_\Phi)$ and $
\tilde \phi  \in H^0(C,L_{\tilde \Phi})$, where $\phi$ is required to be non-vanishing.
\end{itemize}
The structure of the moduli space and the effective supersymmetric quantum mechanics depends intricately on the flux $\fm$ and background line bundles $L_A$ and $L_T$. We will first consider the case $g > 0$ choosing for simplicity to turn off background line bundles for flavour symmetries, before returning to perform a more systematic analysis for $g=0$ in the presence of non-vanishing background fluxes.

\subsubsection{Genus $g>0$}

\begin{figure}[htp]
\centering
\includegraphics[height=2.25cm]{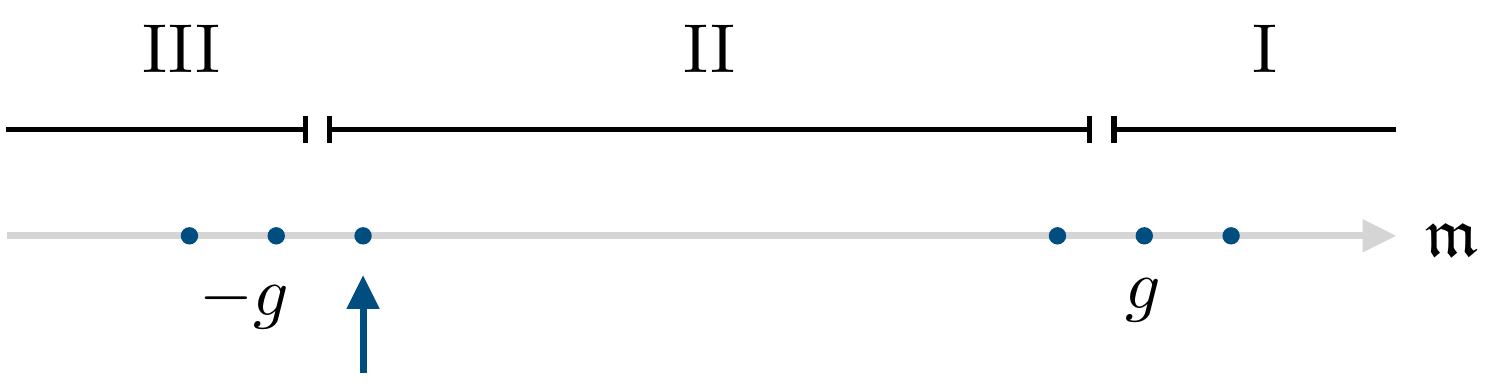}
\caption{We separate our computation of the supersymmetric vacua for supersymmetric QED into three different regions depending on the flux $\fm \in \mathbb{Z}$. An arrow marks point in region II discussed in the main text.}
\label{fig:sqed-regions}
\end{figure}

Let us then consider $g>0$ with trivial background line bundles $L_T = L_A = \cO_C$. The structure of the supersymmetric quantum mechanics depends on the flux $\fm$. We separate our discussion into the three regions shown in figure~\ref{fig:sqed-regions}.

Region I corresponds to $\fm \geq g$. In this region,
\bea
h^0(L_\Phi) & = \fm \qquad && h^0(L_{\tilde\Phi}) = 0 \\
h^1(L_\Phi) & = 0 \qquad  && h^1(L_{\tilde\Phi}) = \fm \, ,
\eea
independent of the line bundle $L$. There are therefore exactly $\fm$ chiral multiplet fluctuations from $\Phi$ and $\fm$ Fermi multiplet fluctuations from $\tilde \Phi$. The underlying moduli space of vortices is again a symmetric product
\be
\cM = \sym^{\fm_\Phi}C \, .
\ee
Since $\fm_\Phi > 2g-2$, the symmetric product is a holomorphic fibration $\cM_\fm \to J_C$ with fiber $\mathbb{CP}^{\fm-1}$. The Fermi multiplet fluctuations from $\tilde\Phi$ then transform as a holomorphic section of the vector bundle $\mathcal{F} = \cO(1) \otimes \mathbb{C}^\fm$ over each fiber.

The existence of the holomorphic fibration implies that the space of supersymmetric ground states factorizes into contributions from the fibre and base,
\be
\cH_\fm = \cH^{(b)}_{\fm} \otimes \cH^{(f)}_{\fm} \, .
\ee
The contribution from the base is
\bea
\cH^{(b)}_{\fm} & = \bigoplus_{q=0}^g H^{0,q}_{\bar\partial}(J_C) \\
& = \sum_{q=0}^g \wedge^q( \C^g ) \, .
\eea
The contribution from the fiber is
\bea
\cH^{(f)}_{\fm} & = \bigoplus_{p=0}^\fm \bigoplus_{\al=1}^{\fm}  H^{0,p}_{\bar\partial}\left(\mathbb{CP}^{\fm-1},K_{\mathbb{CP}^{\fm-1}}^{1/2}\otimes  \frac{\wedge^\al\cF}{\sqrt{\det \cF}} \right)  \\
& = \bigoplus_{p=0}^\fm \bigoplus_{\al=1}^{\fm}  H^{0,p}_{\bar\partial}\left(\mathbb{CP}^{\fm-1},  \cO(\al-\fm) \right) \otimes \wedge^\al(\C^\fm) \\
& = H^{0,0}_{\bar\partial}(\mathbb{CP}^{m-1},\cO) \oplus H^{0,\fm-1}_{\bar\partial}(\mathbb{CP}^{m-1},\cO(-\fm))
\eea
Putting these contributions together and introducing parameters $q$ and $y$ to keep track of the $U(1)_T$ and $U(1)_A$ symmetries respectively, we find
\be
\cH_\fm = (-1)^\fm q^\fm (y^\fm \C - y^{-\fm}\C)  \otimes  \wedge^\bullet( \C^g )\, .
\label{eq:SQED.Hilb.g>0.1}
\ee
In section~\ref{sec:sqed-xyz}, we will demonstrate that this result is in complete agreement with 3d mirror symmetry.

An immediate consequence of the factorization into contributions from the fiber and base is that the supersymmetric twisted index should vanish in region I due to the complete cancellation
\be
\wedge^\bullet( \C^g ) \longrightarrow \sum_{j=0}^g (-1)^j{g \choose j} = 0 
\ee
for $g>0$. This is consistent with the supersymmetric localization result~\eqref{eq:sqed-index}, which for $\fm_T = \fm_A = 0$ is a finite Laurent polynomial in $q$ with maximum power $q^{g-1}$.

Region II corresponds to $-g < \fm < g$. While
\bea
h^0(L_\Phi) - h^1(L_\Phi) & = \fm \\
h^0(L_{\tilde\Phi}) - h^1(L_{\tilde\Phi}) & = - \fm \, ,
\eea
the numbers of chiral and Fermi multiplet fluctuations may jump as the holomorphic line bundle $L$ varies over the Jacobian $J_C$. The computation of the supersymmetric vacua is more difficult in the region and we do not present a general answer.

An exception is $\fm = 1-g$, which is marked with an arrow in figure~\ref{fig:sqed-regions}. In this case, the holomorphic line bundle $L_\Phi$ has degree $\fm_\Phi = 0$. The stability condition requires the existence of a non-zero holomorphic section $\phi$, so we conclude that
\be
L_\Phi =\cO_C \qquad L_{\tilde\Phi} = K_C 
\ee
corresponding to $L = K_C^{-1/2}$. Recall that there is a holomorphic map $\cM_\fm \to J_C$ to the Jacobian parametrizing $L$. In this case, the map is particularly simple: there is one non-vanishing fiber over a single point in the Jacobian. This fiber is parametrized by chiral multiplet fluctuations of the meson $M = \Phi\tilde\Phi$ valued in $H^0(K_C)$ and the moduli space is therefore $\cM_\fm = \C^g$. In addition there is a Fermi multiplet fluctuation $\eta \in H^1(K_C)$. 

Since the moduli space is non-compact, it is essential to turn on a real mass parameter $m_A$ for the $U(1)_A$ axial flavour symmetry. We will choose $m_A>0$. Noting that the meson fluctuations have charge $+2$ under the axial flavour symmetry, we conclude that the twisted Hilbert space for $\fm = 1-g$ is
\be
\cH_{(1-g)} = q^{1-g} y^{g-1}(1-y^2) \bigoplus_{n \geq 0} y^{2n} S^n( \C^g  )   \, .
\label{eq:SQED.Hilb.g>0.2}
\ee
The contribution to the supersymmetric twisted index is therefore
\bea
I_{1-g} & = q^{1-g} y^{g-1} (1-y^2) \sum_{n\geq 0} y^{2n} { n+g-1 \choose g-1} \\
& = q^{1-g}\left(\frac{y}{1-y^2}\right)^{g-1} \, .
\eea
This agrees with the result from supersymmetric localization~\eqref{eq:sqed-index} for $\fm_T = \fm_A = 0$.

Finally, region III corresponds to $\fm \leq-g$. In this region, $h^0(L_\Phi) = 0$. This is incompatible with the stability condition and therefore there are no supersymmetric ground states in this region. The result is consistent with the localization expression~\eqref{eq:sqed-index}, which for $\fm_T = \fm_A = 0$ is a finite Laurent polynomial with minimum power $\zeta^{1-g}$.

\subsubsection{Genus $g=0$}

\begin{figure}[htp]
\centering
\includegraphics[height=8cm]{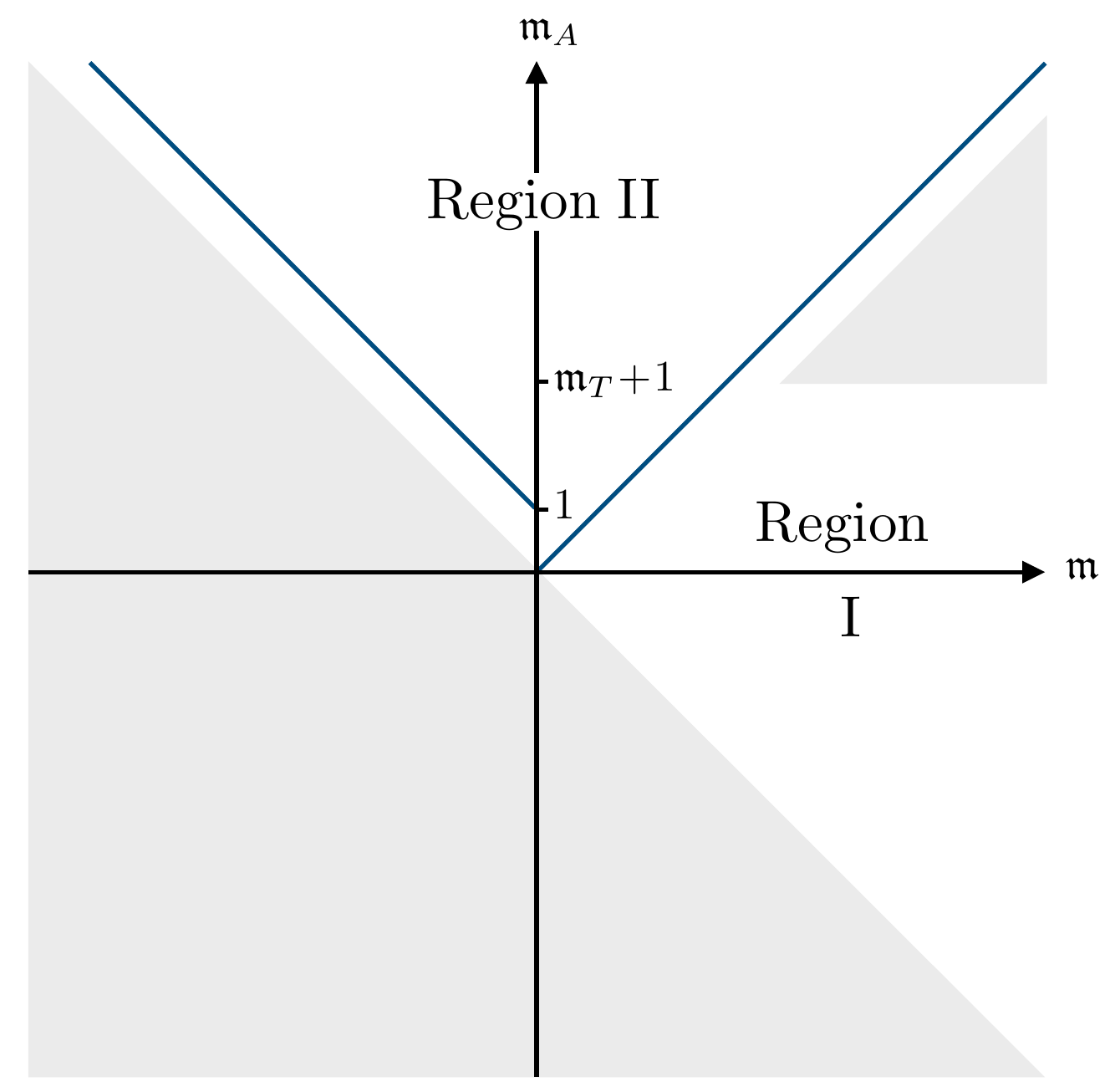}
\caption{Summary of the structure of the moduli space and supersymmetric ground states in the $(\fm,\fm_A)$ plane. There are no supersymmetric ground states in the shaded regions.}
\label{fig:4_6_3}
\end{figure}

For genus zero, we perform a more systematic analysis with non-vanishing background fluxes $\fm_A$, $\fm_T$ for the flavour symmetries. The holomorphic line bundles on $C = \mathbb{CP}^1$ associated to the chiral multiplets are
\be
L_\Phi := \cO(\fm+\fm_A-1) \qquad L_{\tilde \Phi} := \cO(-\fm+\fm_A-1)\, .
\ee
To guide the reader through our analysis, we summarize the structure of moduli spaces $\cM_\fm$ and corresponding contributions to the twisted Hilbert space in the $(\fm,\fm_A)$ plane in figure~\ref{fig:4_6_3}.

First, we note that in order to satisfy the stability condition we require that $h^0(L_\Phi) >0$. This immediately implies that the moduli space is empty and there are no supersymmetric ground states in the region $\fm +\fm_A \leq 0$. This corresponds to the large shaded region in figure~\ref{fig:4_6_3}.

Now assuming that $\fm+\fm_A >0$, the structure of the moduli space $\cM_\fm$ depends critically on the combination $-\fm+\fm_A$, which determines whether $\tilde \Phi$ generates chiral or Fermi multiplet fluctuations in the supersymmetric quantum mechanics. We therefore consider two distinct regions separated by the line $\fm = \fm_A$, as shown in figure~\ref{fig:4_6_3}.

\subsubsection*{Region I}

We first consider $-\fm+\fm_A \leq 0$. This region is characterised by chiral multiplet fluctuations from $\Phi$ and Fermi multiplet fluctuations from $\tilde \Phi$:
\begin{itemize}
\item{$n_{C,\Phi} = \fm + \fm_A > 0$ .}
\item{$n_{F,\tilde\Phi}= \fm - \fm_A \geq 0$ .}
\item{$n_{F,\Phi}= n_{C, \tilde \Phi}=0$ .}
\end{itemize}
The moduli space of vortices is therefore parametrized by the non-vanishing chiral multiplet fluctuations of $\Phi$ modulo complex rescaling,
\be
\mathcal{M}_\fm = \mathbb{CP}^{\fm + \fm_A-1}.
\ee
The Fermi multiplet fluctuations $\bar\eta_{\tilde\Phi}$ have gauge charge $+1$ and therefore transform as sections of the holomorphic vector bundle $\mathcal{F}:= \mathcal{O}(1)\otimes \mathbb{C}^{\fm-\fm_A}$ over $\cM_\fm$.

The contribution to the twisted Hilbert space is therefore given by~\footnote{This computation can be encoded in the cohomology of superprojective space $\mathbb{CP}^{\fm + \fm_A-1|\fm-\fm_A-1}$.
} 
\be
\cH_\fm = H^{0,\bullet}_{\bar\partial} \left(\cM_\fm , K^{1/2}_{\cM_\fm} \otimes\frac{ \wedge^\bullet \cF}{\sqrt{\det(\cF)} } \otimes \widetilde {L}_T\right) \,
\ee
where $\widetilde L_T$ is the holomorphic line bundle on the moduli space induced by the holomorphic line bundle $L_T = \cO(\fm_T)$ on the curve associated to the topological symmetry. First, we notice that the combination 
\be
K^{1/2}_{\cM_\fm} \otimes \frac{1}{\sqrt{\det(\cF)}} = \cO(-\fm) \, .
\ee 
Second, the same argument as section~\ref{sec:U(1)1/2} shows that $\widetilde L_T = \mathcal O (\fm_T)$. Assembling the various pieces, we find
\bea
\cH_\fm & = H^\bullet \left(\cM_\fm , \cO(\fm_T-\fm) \otimes \wedge^\bullet \cF \right)  \\
& = \bigoplus_{i=0}^{\fm-\fm_A} H^\bullet \left( \mathbb{CP}^{\fm+\fm_A-1} , \cO(\fm_T-\fm+i) \right) \otimes  \wedge^i ( \C^{\fm-\fm_A} ) \, .
\eea
Including the flavour symmetry and Fermion number grading, we find
\bea
\cH_\fm = 
&   q^\fm y^{\fm}  \bigoplus_{i=0}^{-\fm_A -\fm_T} S^{-\fm_T -\fm_A - i } (y \mathbb{C}^{\fm+\fm_A})\otimes \wedge^{i} ( -y^{-1} \mathbb{C}^{\fm-\fm_A}) \\
& \oplus (-1)^{\fm+\fm_A-1} \,q^\fm y^{-\fm_A} \bigoplus_{\max \{0,-\fm_T+ \fm\}}^{\fm-\fm_A} S^{-\fm + \fm_T +i} (y^{-1} \mathbb{C}^{\fm+\fm_A})\otimes \wedge^{i} ( -y^{-1} \mathbb{C}^{\fm-\fm_A}) 
 \,,
\label{eq:Hilb.SQED.g=0.R-I}
\eea
where the first line arises from the degree zero cohomology and the second from the maximum degree $\fm+\fm_A-1$ cohomology. If follows from the above formula that there are no supersymmetric ground states when both $\fm > \fm_A$ and $\fm_A > \fm_T$. This is the shaded part of region I on the right of Figure \ref{fig:4_6_3}.

The contribution to the supersymmetric twisted index is therefore
\bea
\mathcal{I}_{\fm} = q^\fm 
 &\sum_{i=0}^{-\fm_T-\fm_A}y^{\fm-\fm_A-\fm_T-2i}(-1)^{i} {{-\fm_T+\fm-1-i}\choose{-\fm_A-\fm_T-i}}{{\fm-\fm_A}\choose i} \\
 &+ q^\fm \sum_{\max \{0, -\fm_T +\fm \}}^{\fm -\fm_A } y^{\fm-\fm_A -\fm_T -2i}(-1)^{\fm+\fm_A-1+i}  {{\fm_A+\fm_T-1+i}\choose{\fm_T-\fm+i}}{{\fm-\fm_A}\choose i} \, ,
 \label{eq:Ind.SQED.g=0.R-I}
\eea
which agrees with the coefficient of $q^\fm$ in the expansion of the supersymmetric twisted index~\eqref{eq:sqed-index} with $g=0$, up to an overall sign $(-1)^{\fm_A}$.

\subsubsection*{Region II}

We now consider the remaining region $\fm_A >|\fm|$. This is characterized by chiral multiplet fluctuations from both $\Phi$ and $\tilde \Phi$. The vortex moduli space is therefore a toric variety given by the total space of the vector bundle
\be
\cM_\fm =  \mathcal{O} (-1)^{\fm_A-\fm} \rightarrow \mathbb{CP}^{\fm+\fm_A-1}  \, .
\label{eq:moduli-line}
\ee
The base is parametrized by the $\fm + \fm_A$ chiral multiplets from $\Phi$ modulo holomorphic gauge transformations and the fiber by the $-\fm+ \fm_A$ chiral multiplets from $\tilde\Phi$. 

This moduli space is non-compact but the $U(1)_A$ flavour symmetry acts on the moduli space $\cM_\fm$ with fixed locus given by the compact zero section $\mathbb{CP}^{\fm+\fm_A-1}$. Therefore, turning on a real mass parameter $m_A$ for the flavour symmetry $U(1)_A$, the effective supersymmetric quantum mechanics is gapped and the contribution to the twisted Hilbert space is
\be
\mathcal{H}_{\fm} = H^{0,\bullet}_{\bar \partial_{m_A}}(\cM_\fm,\sqrt{K_{\cM_{\fm}}} \otimes \tilde L_T) \, .
\label{eq:SQEDg0HilbRegionII}
\ee
where $\bar \partial_{m_A}$ is the Dolbeault operator deformed by the real mass parameter $m_A$ and $L^2$-cohomology classes are understood. 

We will not attempt a direct computation here, except in the special case $\fm = -\fm_A + 1$ corresponding to the blue line in figure~\ref{fig:4_6_3}. Here, the base of the fibration in equation~\eqref{eq:moduli-line} collapses to a point and the moduli space consists of the fiber $\cM_{\fm}  = \bC^{2 \fm_A-1}$ parametrized by fluctuations the meson $M = \Phi\tilde\Phi$. Quantizing these fluctuations, we find
\be
\mathcal{H}_{\fm} =  y^{2\fm_A-1-\fm_T} \bigoplus_{j=0}^\infty S^{j} \left( y^2 \bC^{2\fm_A-1} \right) \, ,
\ee
where $y^2$ is the weight of each coordinate on $\cM_{\fm}  = \bC^{2 \fm_A-1}$ and the overall contribution $y^{2\fm_A-1-\fm_T}$ is the weight of the Fock vacuum induced by $K_{\cM_{\fm_A}} \otimes \widetilde L_T$. This agrees with the contribution to the supersymmetric twisted index
\be
\mathcal{I}_{\fm} = (-1)^{1 + \fm_A} y^{-\fm_T}\left( \frac{y}{1-y^2} \right)^{2\fm_A-1}\, ,
\label{eq:Ind.SQED.g=0.R-II_a}
\ee
modulo a sign $(-1)^{\fm_A+1}$.

We finally provide two complete examples for fixed background fluxes $(\fm_A,\fm_T)$ by summing the contributions from the entire range of $\fm$. 
\begin{itemize}
\item $\fm_A=0$, $\fm_T = 0$ \\
There are no supersymmetric ground states with $\fm \leq 0$. Summing the contributions from $\fm>0$, it follows from eq.~\eqref{eq:Hilb.SQED.g=0.R-I} that
\be
\cH =   \bigoplus_{\fm = 1}^\infty q^\fm ( y^{\fm} \C - y^{-\fm} \C ) 
\ee
with supersymmetric twisted index
\be
\mathcal{I} =  \sum_{\fm = 1}^\infty q^\fm (y^{\fm} - y^{-\fm}) = \frac{1-y^2}{y^2 \left(1-\frac{1}{qy}\right) \left(1-\frac{q}{y}\right)} \, .
\ee
\item $\fm_A=1$, $\fm_T=0$ \\
There are no supersymmetric ground states with $\fm < 0$ as the moduli space is empty. For $\fm>0$, the moduli space is $\cM_\fm = \mathbb{CP}^\fm$ but nevertheless the contribution~\eqref{eq:Ind.SQED.g=0.R-I} to the twisted Hilbert space vanishes. Finally, the non-vanishing contribution from $\fm = 0$ gives 
\be
\cH = y \, S^\bullet (y^2\C)
\ee
with supersymmetric twisted index
\be
\mathcal{I}=\frac{y}{1-y^2}\, .
\ee 
\end{itemize}


\section{Mirror Symmetry}
\label{sec:mirror}

In this section, we will compare results obtained in section~\ref{sec:examples} to provided new checks of two basic instances of 3d mirror symmetry.


\subsection{Particle-Vortex Duality}
\label{sec:U(1)-chiral-mirror}

We consider the following pair:
\begin{itemize}
\item I. Chiral multiplet with $R$-charge $r=0$ and supersymmetric Chern-Simons terms $k_{ff} = k_{fR} = -\tfrac{1}{2}$ for the $U(1)_f$ flavour symmetry.
\item II. $U(1)$ Chern-Simons theory at level $k=+\frac{1}{2}$ with one chiral multiplet of charge $+1$ and $R$-charge $r=1$ .
\end{itemize}
For positive FI parameter $\zeta>0$ in theory II, the monopole operator of topological charge $-1$ is gauge neutral and can be identified with the chiral multiplet in theory I. The $U(1)_f$ flavour symmetry of theory I is identified with the $U(1)_T$ topological symmetry of theory II and therefore we make the identifications $m_f = \zeta$ and $x = q$.

We first compare the supersymmetric twisted Hilbert spaces for $g\geq 1$. Let us first recall the supersymmetric twisted Hilbert space of theory I with a background line bundle $L_f$ for the $U(1)_f$ flavour symmetry in the chamber $\mathfrak{c}_+= \{m_f>0\}$. From equation~\eqref{eq:ChiralHilbertSpace} we find
\bea
\mathcal{H}^{(\mathrm{I})}_+ & = x^{-\frac{\fm_f}{2}} x^{-\frac{g-1}{2}} \cdot x^{\frac{\fm_f-g+1}{2}} \bigoplus_{j=0}^\infty x^j\left( \bigoplus_{p+q=j} S^pH^0(L_f) \otimes \wedge^q H^1(L_f) \right) \\
& = x^{1-g} \bigoplus_{j=0}^\infty x^j\left( \bigoplus_{p+q=j} S^pH^0(L_f) \otimes \wedge^q H^1(L_f) \right) \, ,
\eea
where $h^0(L_f)-h^1(L_f) = \fm_f -g +1$. The additional factor $x^{-\frac{\fm_f}{2}} x^{-\frac{g-1}{2}}$ arises from the shift in the charge of the Fock vacuum induced by the mixed supersymmetric Chern-Simons terms $k_{ff} = k_{fR} = -\tfrac{1}{2}$. 

Now recall the supersymmetric twisted Hilbert space of theory II in the chamber $\mathfrak{c_+} = \{\zeta >0\}$ for the FI parameter. Introducing a background line bundle $L_T$ for the topological symmetry, from equation~\eqref{eq:U(1)1/2HilbertSpace} we find
\bea
\cH_+^{(\mathrm{II})} & = \bigoplus_{\fm \geq 1-g} q^{\fm} \left( \bigoplus_{p+q= \fm+g-1} S^p H^0(L_T) \otimes \wedge^q H^1(L_T) \right) \\
 & = q^{1-g} \bigoplus_{j=0}^\infty q^j\left( \bigoplus_{p+q=j} S^pH^0(L_T) \otimes \wedge^q H^1(L_T) \right)
\eea
where $h^0(L_T) - h^1(L_T) = \fm_T-g+1$. Under the identifications $m_f = \zeta$, $x = q$ and $L_f = L_T$, we find complete agreement between the supersymmetric twisted Hilbert spaces. 

This constitutes a more refined check of mirror symmetry than the supersymmetric index. First, $\cH$ contains an infinite number of supersymmetric ground states whereas the index $\cI$ truncates to a finite Laurent polynomial. Second, we find agreement between the supersymmetric Hilbert space as we vary the background line bundle $L_f = L_T$ over the Jacobian $J_C$.
The index cannot detect this phenomenon, as it depends only on the degree of $\fm_f = \fm_T$.

Let us briefly consider the genus zero case, including the parameter $\xi$ for the additional flavour symmetry $U(1)_\ep$ rotating $C = \mathbb{CP}^1$. For simplicity, we assume that $\fm \geq0$. The supersymmetric twisted Hilbert spaces are
\bea
\cH^{(\mathrm{I})}_+ & = \bigoplus_{j=1}^\infty x^j S^{j-1}(\xi^\rho\C^{\fm+1}) \\
\cH^{(\mathrm{II})}_+ & = \bigoplus_{j=1}^\infty x^j S^{\fm}(\xi^\rho\C^{j}) \, ,
\eea
which agree due to the following isomorphism of graded vector spaces,
\be
S^{j-1}(\xi^\rho\C^{\fm+1}) \cong S^{\fm}(\xi^\rho\C^{j}) \, .
\ee

\subsection{XYZ  $\leftrightarrow$ SQED}
\label{sec:sqed-xyz}

We now turn to the investigation of the following pair:
\begin{itemize}
\item I. Three chiral multiplets with cubic superpotential $W = XYZ$ and charge matrix:
\begin{center}
\begin{tabular}{c | c c c }
 & $U(1)_T$ & $U(1)_A$ & $U(1)_R$ \\
\hline
$X$  & $1$ & $-1$ & $0$ \\
$Y$  & $-1$ & $-1$ & $0$\\
$Z$ & $0$ & $2$ & $2$
\end{tabular} .
\end{center}
\item II. $U(1)$ supersymmetric gauge theory with chiral multiplets $\Phi$, $\tilde\Phi$ with charge matrix:
\begin{center}
\begin{tabular}{c | c c c c}
& $U(1)$ & $U(1)_T$ & $U(1)_A$ & $U(1)_R$ \\
\hline
$\Phi$ & $1$ & $0$ & $1$ & $1$ \\
\hline
$\tilde \Phi$ & $-1$ & $0$ & $1$ & $1$ 
\end{tabular} .
\end{center}
\end{itemize}
Here we identify from the outset the $U(1)_T \times U(1)_A$ flavour symmetry. Under 3d mirror symmetry, $Z$ is mapped to the gauge invariant combination $\Phi \tilde\Phi$ whereas $X$,$Y$ are mapped to monopole operators in the supersymmetric gauge theory of charge $\pm1$ under the topological symmetry.

Let us first consider $g>0$. We will not attempt an exhaustive analysis but specialize to $L_A = L_T = \cO_C$ and real mass parameters in the chamber 
\be
\mathfrak{c}_{+-+} = \{ 0 < m_A < m_T\} 
\ee
where have have obtained explicit results on both sides. This corresponds to an expansion of the supersymmetric twisted index in the region $|q| < |y| < 1 $. 

First, there are no supersymmetric ground states of topological charge $q_T \leq g$ on both sides. Second, we consider the supersymmetric ground states of charge $q_T \geq g$, whose the contributions to the supersymmetric index vanish. Comparing equations~\eqref{eq:Hilb.XYZ.g>0.1} and~\eqref{eq:SQED.Hilb.g>0.1} we find the contribution to $\cH_{+-+}$ in both theories is
\bea
q^{q_T} ( y^{q_T}\C - y^{-q_T}\C ) \otimes  \wedge^\bullet ( \mathbb{C}^g ) \qquad q_T \geq g
\eea
modulo an overall sign. Note that $q_T = \fm$. Finally, in the intermediate region $-g < q_T < g$ we have only computed explicitly the supersymmetric ground states with minumum charge $q_T = 1-g$. Comparing equations~\eqref{eq:Hilb.XYZ.g>0.2} and~\eqref{eq:SQED.Hilb.g>0.2}, we find that the contribution to $\cH_{+-+}$ in both theories is
\be
q^{1-g} y^{g-1}(\C-y^2\C) \otimes \bigoplus_{n = 0}^\infty y^{2n} S^n(\C^g)  \, .
\ee
In regions where we can independently compute the supersymmetric ground states, we therefore find agreement with mirror symmetry.

We now study the case $g=0$ with background fluxes $\fm_T$, $\fm_A$ for the flavour symmetry. We separate the analysis into characteristic regions in the $(\fm_T,\fm_A)$ plane:
\begin{itemize}
\item In the region $\fm_A \leq 0$, $ \fm_T \geq -\fm_A$ we have computed the supersymmetric twisted Hilbert space in both theory I and II. These computations agree, but there are no cancellations in passing to the supersymmetric twisted index in this case, so we cannot provide a stronger check of mirror symmetry. 
\item In the region $\fm_A\leq 0$, $0 \leq \fm_T< -\fm_A$, we have computed the supersymmetric twisted Hilbert space of theory II in equation~\eqref{eq:Hilb.SQED.g=0.R-I}. In this case, there are cancellations between bosons and fermions in computing the supersymmetric twisted index. However, we have not performed an independent computation for theory I. Nevertheless, it is possible to check that the supersymmetric ground states in~\eqref{eq:Hilb.SQED.g=0.R-I} form a subset of those for the three chiral multiplets $X$, $Y$, $Z$ in the absence of a superpotential. This is consistent with the presence of a non-vanishing $J$-term differential in supersymmetric quantum mechanics for theory I in this region and equation~\eqref{eq:Hilb.SQED.g=0.R-I} can be regarded as a prediction for its cohomology.
\item In the other direction, we can use our result for supersymmetric ground states of theory I to make a prediction for the supersymmetric ground states in the unknown region $\fm_A \geq 1$, $|\fm_T| < \fm_A$ of theory II. From equation~\eqref{eq:XYZg0Hilb1},
\bea
\mathcal{H}^{(I)}_{+-+} = &q^{\fm_T} y^{3\fm_A-2} S^{\bullet} (y^2 \mathbb{C}^{2 \fm_A-1}) \otimes \\ &\otimes \bigoplus^{-\fm_T+\fm_A-1}_{\fm=-\fm_T-\fm_A+1} q^{\fm} \left( \bigoplus_{i-k=\fm} \wedge^{i} (y^{-1} \mathbb{C}^{-\fm_T+\fm_A-1}) \otimes \wedge^{k} (y^{-1} \mathbb{C}^{\fm_T+\fm_A-1}) \right) \, .
\label{eq:XYZg0Hilb1Repetition}
\eea
This should reproduce the cohomology expressed in equation~\eqref{eq:SQEDg0HilbRegionII},
\be
\mathcal{H}^{(II)}_{+-+} =   H_{\bar{\partial}_{m_A}}^{0,\bullet} ( \cM_\fm , \sqrt{K_{\cM_{\fm}}} \otimes \widetilde L_T ) \, ,
\ee
where
\be
\cM_\fm = \mathcal{O} (-1)^{\fm_A-\fm} \rightarrow \mathbb{CP}^{\fm+\fm_A-1} \, .
\ee
In the degenerate case $\fm = -\fm_A + 1$ where the base collapses to a point, equation~\eqref{eq:XYZg0Hilb1Repetition} correctly reproduces the cohomology of the unique fibre $\cM_\fm = \bC^{2\fm_A-1}$. It would be interesting to understand how to compute this cohomology in the general case.
\end{itemize}

Finally, it is straightforward to demonstrate agreement between regions where there are no supersymmetric ground states. Taking this into consideration, we are able to chart the supersymmetric ground states on almost the whole range of parameters $(\fm_T,\fm_A,\fm)$, aside from a small region depending on $\fm_T$.


\section{Discussion}
\label{sec:disc}

We conclude with some discussion of our results and directions for further research.

We considered here only the simplest abelian gauge theories with $G = U(1)$ and an FI parameter $\zeta$ such that the space supersymmetric ground states was captured by an effective sigma model onto moduli spaces of `Higgs branch' vortex equations. However, our analysis was not quite complete as there may exist chambers where there are `topological' solutions that must be accounted for. For example, this happens in the example of section~\ref{sec:U(1)1/2} for a $U(1)_{\frac{1}{2}}$ gauge theory with a chiral multiplet of charge $+1$ in the chamber $\zeta <0$. The presence of such topological solutions is indicated in the supersymmetric twisted index by the presence of residues at infinity in the contour integral expressions of~\cite{Benini:2015noa,Benini:2016hjo,Closset:2016arn}. This will be addressed in forthcoming work~\cite{line-op}.

More generally, for non-abelian $G$ we expect to encounter singular moduli spaces with loci where the gauge symmetry is not completely broken. In such cases, an effective supersymmetric sigma model is unlikely to capture the supersymmetric ground states exactly. The relevant moduli spaces are better described as quotient stacks, indicating that aspects of the full gauge theory description will be needed to compute the supersymmetric ground states. Nevertheless, we hope at least the supersymmetric twisted index can be understood as a generating function of virtual Euler characters of moduli stacks of generalized vortices or `quasi-maps' as in reference~\cite{Okounkov:2015spn}.

The focus in this paper has been on understanding the supersymmetric ground states of the effective supersymmetric quantum mechanics on $\R$. It would be interesting to consider the spectrum of operators in the supersymmetric quantum mechanics in the cohomology of the supercharge $\bar Q$ and their matrix elements, along the lines of \cite{Bullimore:2016hdc}. Such operators will arise from local operators in three dimensions, such as elementary scalar fields and monopole operators in chiral multiplets, but also from supersymmetric line operators wrapping cycles on $C$. 

The construction of the supersymmetric ground states presented here can be extended by adding supersymmetric line operators preserving the 1d $\cN=(0,2)$ supersymmetry algebra on $\R$. This leads to a modification of the supersymmetric twisted Hilbert space, which will be addressed in a separate publication~\cite{hilb-top}. Examples include supersymmetric Wilson lines and coupling to chiral and Fermi multiplets via superpotentials. This should lead to more refined checks of 3d mirror symmetry for line operators~\cite{Assel:2015oxa}.

Boundary conditions preserving 2d $\cN=(0,2)$ supersymmetry have been studied in~\cite{Gadde:2013sca,Okazaki:2014lda,Dimofte:2017tpi}. In the present setup, truncating to $\R_+$ with such a supersymmetric boundary condition wrapping the Riemann surface $C$ will preserve a common supercharge $\bar Q$. We therefore expect the boundary condition to define a cohomology class in the supersymmetric twisted Hilbert space (for related constructions with more supersymmetry see~\cite{Bullimore:2016nji,Bullimore:2016hdc}). This should allow for new checks of dual pairs boundary conditions introduced in~\cite{Dimofte:2017tpi}.

There is an $SL(2,\mathbb{Z})$ action on 3d theories with a $U(1)_f$ flavour symmetry generated by adding a Chern-Simons term at level $+1$ and gauging the flavour symmetry~\cite{Witten:2003ya}. It would be interesting to understand such an action on the family of supersymmetric twisted Hilbert spaces parametrized by a choice of holomorphic line bundle $L_f$ on $C$. Adding a supersymmetric Chern-Simons term corresponds simply to a shift of the flavour grading. It would therefore be necessary to understand `gauging' as an algebraic operation at the level of the supersymmetric twisted Hilbert space.

The $SL(2,\mathbb{Z})$ action mentioned above plays an important role in the 3d-3d correspondence~\cite{Dimofte:2011py,Dimofte:2011ju,Cecotti:2011iy}. It would be particularly interesting to construct a version of this correspondence for the supersymmetric twisted Hilbert space starting from the 6d $\cN=(0,2)$ theory on $\mathbb{R} \times C \times M_3$. In the first instance, this should lead to a correspondence between:
\begin{itemize}
\item The twisted Hilbert space of the 3d $\cN=2$ theory $T(M_3)$ on $\R \times C$.
\item The Hilbert space of the Donaldson-Witten twist of the 4d $\cN=2$ theory $T(C)$ on $\R \times M_3$.
\end{itemize}
This dictionary could then be extended to include the presence of various extended defects.
The investigation of this correspondence has been initiated in~\cite{Gukov:2017kmk,Gukov:2016gkn} and we hope the results presented here will contribute to further progress in this area.


\section*{Acknowledgements}

We would like to thank Chris Beem, Tudor Dimofte, Jacques Distler, Andre Henriques, Heeyeon Kim, Tomasz Lukowski for useful discussions related to this project. We would also like to thank the organizers and participants of the Pollica Summer Workshop on Dualities in Superconformal Field Theories for providing a stimulating environment where part of this work was completed. MB gratefully acknowledges partial support from ERC STG grant 306260. AF gratefully acknowledges support from Janggen-Poehn Stiftung, St. Gallen.


\appendix

\section{Mathematical Background}
\label{app:vortex-symmetric}

A prototypical example for the target space of the supersymmetric quantum mechanics we are considering is the moduli space of abelian vortices on a Riemann surface $C$. There is an extensive literature on the topic \cite{Jaffe:1980mj,GarciaPrada:1993qv,Manton:2004tk}, and the goal of this appendix is to summarize some elements needed for this paper.

\subsection{Vortices and Symmetric Products}
\label{app:vortices}

Let $L$ be a hermitian line bundle of degree $d$ on a Riemann surface $C$. We consider a smooth unitary connection $A$ on $L$ and a smooth section $\phi$.  Let $\mathcal{V}_d$ denote the space of pairs $(A,\phi)$ that are solutions to the vortex equations on $C$,
\bea
\frac{1}{e^2} *F_A + \la \phi , \phi\ra= \zeta \\
  \bar\partial_A \phi = 0 \, ,
  \label{eq:app-vortex}
\eea
where $F_A$ is the curvature of the connection $A$ and $\bar\partial_A$ is the holomorphic structure on $L$ inherited from $d_A$ and the complex structure on $C$. Furthermore, let $\mathcal{G}$ be the group of gauge transformations, $\mathcal{G}= \text{Hom} (C, U(1))$. 
The moduli space of vortices is defined by the quotient 
\be
\mathcal{M}_d := \mathcal{V}_{d}\, /\, \mathcal{G} \, .
\ee
By integrating the first vortex equation in~\eqref{eq:app-vortex}, we learn the moduli space can only be non-empty if
\be
\zeta \geq \frac{2\pi d}{e^2\text{Vol}_C} \, .
\ee
We assume the strict version of this inequality in what follows.

The moduli space of vortices can be understood as an infinite-dimensional K\"ahler quotient. First, the space of pairs $(A,\phi)$ is an infinite-dimensional K\"ahler manifold with flat metric
\be
g = \frac{1}{4\pi}\int_C \left( \frac{1}{e^2}\delta A \wedge * \delta A + * \la \delta \phi , \delta \phi \ra \right)
\ee
inherited from the metric on $C$ and the hermitian metric $\la \, \cdot \, , \, \cdot \, \ra$ on the line bundle $L$. The second vortex equation $\bar\partial_A \phi = 0$ defines a K\"ahler submanifold $\cN_d$ of this space, on which gauge transformations act with moment map
\be
\frac{1}{e^2} *F_A + \la \phi , \phi \ra \, .
\ee
We can therefore express the vortex moduli space as an infinite-dimensional K\"ahler quotient $\cM_d  = \cN_d \, /\!/ \, \cG$.

In our computations, we will not require the K\"ahler metric on $\cM_d$ inherited from the quotient construction. Instead, it is convenient to introduce an algebraic description of the moduli space by replacing the moment map constraint by the stability condition that $\phi$ cannot vanish identically and dividing by complex gauge transformations. 

The moduli space then parametrizes pairs $(L,\phi)$, where $L$ is a holomorphic line bundle of degree $d$ and $\phi$ is a non-vanishing holomorphic section of $L$. Equivalently, the moduli space parametrizes effective divisors 
\be
D = p_1+ \ldots+ p_d \, ,
\ee
such that $L = \cO_C(D)$ and the holomorphic section $\phi$ vanishes on $D$. We therefore have a symmetric product
\be
\cM_d \cong \text{Sym}^dC \, .
\ee
The points $p_1,\ldots,p_d$ are the positions of the vortex centres. 

There is a holomorphic map
\bea
j & : \sym^dC  \to \mathrm{Pic}^d(C)\cong J_C \\
 & : (L,\phi) \mapsto L \\ 
& : \{D\} \mapsto \cO_C(D) \, .
\eea
Provided $n \geq 2g-1$, this is a holomorphic fibration with fiber $\mathbb{CP}^{n-g}$ given by the projective space of global sections $\mathbb{P} H^0(C,L)$, or equivalently the complete linear system associated to $D = p_1 + \cdots + p_n$. For $n< 2g-1$, the structure of this map is studied in Brill-Noether theory.

\subsection{Cohomology of Symmetric Products} 
\label{app:coh}

Our construction of the space of supersymmetric vacua involves computing the (Dolbeault) cohomology of the moduli space of vortices. The simplest way to understand the cohomology of symmetric products is via the isomorphism
\be
H^{\bullet} (\text{Sym}^{d}(C),\mathbb{K}) \cong H^{\bullet} (C^{d},\mathbb{K})^{S_d},
\ee
where $\mathbb{K}$ is any field. The right-hand side consists of permutation invariant elements in the cohomology of the $d$-fold product of $C$. 

Let us introduce standard generators for the cohomology ring of $C$,
\be
\gamma_i  \in H^{1,0}(C,\mathbb{K}) , \quad \tilde{\gamma}_i  \in H^{0,1}(C,\mathbb{K}), \quad \beta  \in H^2(C,\mathbb{K}) \, ,
\ee
where $i=1,\ldots,g$.
They induce cohomology classes in the $d$-fold product of $C$,
\bea
\gamma_{i,j} &= 1 \otimes \cdots \otimes 1 \otimes \gamma_i \otimes 1 \otimes \cdots \otimes 1 \in  H^{1,0}(C^d,\mathbb{K})  \\
\tilde{\gamma}_{i,j} &= 1 \otimes \cdots \otimes 1 \otimes \tilde{\gamma}_i \otimes 1 \otimes \cdots \otimes 1 \in H^{0,1}(C^d,\mathbb{K}) \\
\beta_j &= 1 \otimes \cdots \otimes 1 \otimes \beta_i \otimes 1 \otimes \cdots \otimes 1 \in H^{1,1}(C^d,\mathbb{K}) \, ,
\eea
where the generator appears in the $j$-th factor. The classes 
\be
\Gamma_i = \sum_{j=1}^d \gamma_{i,j} , \quad \tilde{\Gamma}_i = \sum_{j=1}^d\gamma_{i,j} , \quad \eta = \sum_{j=1}^d \beta_j
\label{eq:symm-prod-generators}
\ee
then descend to $H^{\bullet} (C^{d},\mathbb{K})^{S_{d}}$, and in fact generate it. As we will not make use of them, we omit the ring relations and refer the reader to~\cite{MACDONALD1962319}.

It is also useful to understand how the cohomology of a symmetric product is induced from the fibration structure when $d \geq 2g-1$. The cohomology of a fibration with compact fibres can be computed using the Serre spectral sequence. In this particular example, the Serre spectral sequence collapses immediately and
\be
H^{\bullet} (\text{Sym}^{d}C,\mathbb{K}) \cong H^{\bullet}(\mathbb{CP}^{d-g}, \mathbb{K} ) \otimes  H^{\bullet} (J_C, \mathbb{K}).
\ee
Since the Jacobian is isomorphic to a $2g$-dimensional torus, its cohomology is an exterior algebra, whose generators are the classes $\Gamma_i$, $\tilde{\Gamma}_i$ introduced above. The cohomology of the fiber $\mathbb{CP}^{d-g}$ is generated by the Chern class of the dual of the tautological bundle, which is identified with the class $\eta$.

\subsection{Universal Constructions}
\label{app:universal}

In relating constructions on $C$ and the symmetric product of $C$, an important r\^ole is played by universal constructions on $\text{Sym}^{d}(C) \times C$.
We first explain the construction for divisors. The subset
\be
\Delta  = \{(D,p)\in \sym^d(C) \times C \ | \ p \in D \}  \,
\ee 
is called the universal effective divisor on $\text{Sym}^{d}(C) \times C$. An immediate consequence of the definition is that $\Delta$ cuts out on $\{D\} \times C$ precisely the divisor $D$ on $C$. Conversely, given a point $p\in C$ the intersection of the universal divisor with $\sym^d C \times \{p\}$ is $\widetilde D_p \times \{p\}$, where the divisor $\widetilde D_p$ is the image of the inclusion
\bea
i_p & \quad : \quad  \sym^{d-1} (C) \hookrightarrow \sym^{d}C \\
i_p & \quad : \quad  D \mapsto D + p \, .
\eea
The divisor $\widetilde D_p$ defines a class in $H^2(\sym^{d}C, \mathbb{Z})$ which coincides with the cohomology class $\eta$ defined in equation~\eqref{eq:symm-prod-generators}.

This universal construction can also be formulated in terms of line bundles. The divisor $\Delta$ defines a universal line bundle $\cL = \cO(\Delta)$ on $\sym^d(C) \times C$, which can also be obtained as follows. Let $\pi_{\mathcal{V}_d} : \mathcal{V}_d \times C \rightarrow C$ 
denote the projection on the second factor. Then the universal line bundle is
\be
\cL  \cong (\pi_{\mathcal{V}_d}^{*} L) /\mathcal{G},
\label{eq:univ-line-bundle}
\ee
where $L=\cO(D)$ and the group of gauge transformations $\mathcal{G}$ acts on both the space of solutions to the vortex solutions $\mathcal{V}_d$ and the line bundle $L$. We now summarize some important properties of the universal line bundle. First, for each point $[A,\phi] \in \mathcal{M}_d$ there is an isomorphism of $U(1)$-bundles
\be
\mathcal{L}|_{[A,\phi]}\cong L \, .
\ee
Second, $\mathcal{L}$ has a natural connection called the universal connection, which we will denote $\mathcal{A}$. It is induced by the connection on 
\be
\pi_{\mathcal{V}_{d}}^{*} L \rightarrow \mathcal{V}_{d} \times C ,
\ee
which is trivial in the $\mathcal{V}_{d}$ directions and tautological in the $C$ directions, namely on $(A,\phi) \times C$ it acts exactly like $A$. However, in order to pass to the quotient in equation~\eqref{eq:univ-line-bundle} and explain further properties of the universal connection $\mathcal{A}$, we first need to introduce some standard constructions in gauge theory.

We first note that the space of solutions $(A,\phi)$ to the vortex equations is naturally a principal $\mathcal{G}$-bundle over the moduli space, $\mathcal{V}_d \rightarrow \mathcal{V}_d /\mathcal{G} = \mathcal{M}_d$. The tangent space can therefore be decomposed as a direct sum
\be
T\mathcal{V}_d  = T \mathcal{V}_{d, \text{vert}} \oplus T \mathcal{V}_{d, \text{hor}} \, ,
\ee
where the vertical subspace $T \mathcal{V}_{d,\text{vert}}$ is canonically defined as the subspace tangent to the fibres of $\mathcal{V}_d \to \cM_d$: they correspond to infinitesimal gauge transformations. However, without additional structure the horizontal subspace $T\mathcal{V}_{d,\text{hor}}$ is not canonical, and such a decomposition is equivalent to a choice of connection one-form $\theta \in \Omega^1 (\mathcal{V}_{d}, \mathfrak{G})$ where $\mathfrak{G}$ is the Lie agebra of $\mathcal{G}$. Given such a connection one-form, the horizontal subspace is defined by 
\be
T_{(A, \phi)} \mathcal{V}_{\text{hor}} = \{ (\dot{A}, \dot{\phi}) \in T_{(A, \phi)}\mathcal{V} \ | \ \theta_{(A, \phi)} (\dot{A}, \dot{\phi}) = 0 \ \},
\ee
where we follow the standard convention of denoting tangent vectors by $(\dot A ,\dot \phi)$. Conversely, any choice of horizontal subspace gives rise to a connection one-form.

The physical connection one-form is determined by Gauss's law constraint. Let us denote the constraint on tangent vectors $(\dot{A},\dot{\phi})$ over a point $ (A,\phi)\in \mathcal{V}_d$ by $G_{(A,\phi)}$. Then the connection one-form is determined by
\be
\theta_{(A,\phi)} (\dot{A},\dot{\phi}) =  \left\{ \alpha \in \mathfrak{G}  \ | \  G_{(A,\phi)} \left( (\dot{A}, \dot{\phi})+ \alpha (\dot{A}, \dot{\phi}) \right) = 0 \right\}  ,
\ee
where
\be
\alpha (\dot{A}, \dot{\phi}) = ( d_A \alpha ,  i \alpha \cdot \phi),
\ee
is the action of an infinitesimal gauge transformation $\alpha$. 

Let us now return to the universal connection $\mathcal{A}$. This is uniquely determined by the action of its curvature $F_\mathcal{A}$ on tangent vectors $(\dot{A},\dot{\phi},u)\in T_{(A,\phi,p)}(\mathcal{\cM}_d\times C)$. This is given by
\bea
&F_{\mathcal{A}} ((0,0,u_1),(0,0,u_2))|_{ (A, \phi, p )} = F_{A}(u_1,u_2) (p)  \\
&F_{\mathcal{A}} ((\dot{A},\dot{\phi},0),(0,0,u))|_{ (A, \phi, p )} = \dot{A}(u) |_p  \\
&F_{\mathcal{A}} (((\dot{A}_1,\dot{\phi}_1),0),((\dot{A}_2,\dot{\phi}_2),0))|_{ (A, \phi, p )} =  F_{\theta} ((\dot{A}_1,\dot{\phi}_1),(\dot{A}_2,\dot{\phi}_2))|_p\, .
\eea
Here we are abusing notation somewhat by conflating tangent vectors to $\mathcal{M}_d$ and $\mathcal{V}_d$. The universal connection is therefore tautological in the $C$ directions and implements Gauss' law constraint in the $\mathcal{M}_{d}$ directions. At the level of cohomology classes,
\bea
&-\frac{1}{2\pi} [F_{\mathcal{A}}]|_{[A,\phi]\times C} = d  \in H^2(C,\mathbb{Z}) = \mathbb{Z} \, ; \\
&-\frac{1}{2\pi} [F_{\mathcal{A}}]|_{\mathcal{M}_{d}\times \{x \} } = \eta \in H^2 (\sym^{d}(C),\mathbb{Z}) \, ,
\label{eq:universalcurvature}
\eea
which follows from the fact that the connection is tautological along $C$ and the restriction of $\mathcal{L}$ to $\mathcal{M}_d$ is the holomorphic line bundle induced by the divisor $\tilde{D}_p$ with associated class $\eta$. 

\subsection{Line Bundles and Deligne Pairing} 

The universal construction provides a natural way to construct a line bundle $\widetilde L_T$ on the moduli space $\cM_d = \sym^{d}C$ starting from a line bundle $L_T$ on $C$. Our notation reflects the fact that $L_T$ corresponds to a vacuum expectation value for a background vectormultiplet for the $U(1)_T$ topological flavour symmetry.

We consider the following diagram
\begin{center}
\begin{tikzcd}[column sep=normal]
& \sym^{d}C \times C \arrow{dl}[swap]{\pi}\arrow{dr}{p} & \\
\sym^{d}C  & & C \, .
\end{tikzcd}
\end{center}
We start from a holomorphic line bundle $L_T$ on $C$. Then there are two natural holomorphic line bundles on $\sym^{d}C \times C$: the pull-back $p^{*}L_T$ and the universal line bundle $\mathcal{L}$. As explained in reference~\cite{Eriksson:2016wra}, we can then produce a holomorphic line bundle $\widetilde L_T$ on $\sym^{d}{C}$ known as the \emph{Deligne pairing}. This is usually denoted by
\begin{equation}
\widetilde L_T = \langle \cL, p^{*} L_T \rangle  \, .
\label{eq:DelignePairing}
\end{equation}
We will not provide the details of this construction here. However, it is important to point out that given connections on $L_T$ and $\mathcal{L}$ the following equality between the curvatures holds,
\begin{equation}
F_{\widetilde A_T } =\frac{1}{2\pi} \int_{C} F_{\mathcal{A}} \wedge F_{A_T} \, .
\label{eq:DeligneCurvature}
\end{equation}
Furthermore, reference~\cite{Eriksson:2016wra} shows that the holomorphic line bundle~\eqref{eq:DelignePairing} is isomorphic to the holomorphic line bundle denoted similarly in section~\ref{sec:U(1)1/2}, which is obtained via the identification $\sym^dC \cong C^d / S_d$.

\section{Background Line Bundles and Electric Impurities}
\label{app:impurities}

In this appendix, we explain how a background line bundle $L_T$ for a topological flavour symmetry can be understood as an electric impurity in the gauge theory. In particular, we show that the curvature of the `dirty connection' introduced in~\cite{Tong:2013iqa} coincides with the curvature of the Deligne pairing~\eqref{eq:DeligneCurvature}.

We introduce local coordinates $\{ X^a\}$ on the moduli space $\cM_d$ and let $A(x,X)$, $\phi (x,X)$ denote a solution of the vortex equations corresponding to the point in the moduli space with local coordinates $X^a$. We also introduce a local coordinate $x$ on $C$. In the notation of Appendix~\ref{app:vortex-symmetric}, this corresponds to a smooth section $s : \cM_d \to \cV_d$ of the principal bundle $\mathcal{V}_d$. We therefore use a shorthand notation $s(X) = (A(x,X),\phi(x,X))$.

Let us now consider a tangent vector
\be 
\frac{\partial}{\partial X^a} \in T_X \mathcal{M}_d \, .
\ee
The push-forward of this tangent vector to $\cV_d$ is given by 
\be
s_* \left( \frac{\partial}{\partial X^a}  \right) = 
\left(\frac{\partial A }{\partial X^a} \biggr\rvert_X , \frac{\partial \phi }{\partial X^a} \biggr\rvert_X \right) \in T_{s(X)} \mathcal{V}_d \, .
\ee
This tangent vector need not be horizontal with respect to the connection one-form $\theta \in \Omega^1 (\mathcal{V}_{d}, \mathfrak{G})$ and therefore we obtain $\mathfrak{G}$-valued functions on $\mathcal{M}_d \times C$, 
\be
\theta_{s(X)} \left(s_* \frac{\partial}{\partial X^a} \biggr\rvert_X \right)   = \alpha_a (x, X) \, .
\ee
It follows that $s^* \theta = \alpha_a (x,X) \mathrm{d}X^a$ is a connection on $\cM_d$ with covariant derivative
\bea
\delta_a A & = \left( \frac{\partial}{\partial X^a} + d \alpha_a \right) A\\
\delta_a \phi & = \left( \frac{\partial}{\partial X^a} +  i \alpha_a \right) \phi \, .
\eea
The universal connection on $\cM_d \times C$ can then be expressed
\be
\cA(X,x) = \alpha_a (x,X) \mathrm{d}X^a + A(X,x) \, .
\label{eq:univ-conn}
\ee

In reference~\cite{Tong:2013iqa} the bosonic part of the action for the effective supersymmetric quantum mechanics of the collective coordinates $\{X^a\}$ is given by
\be
S = \int d\tau \ g_{ab} (X) \frac{dX^a}{d\tau} \frac{dX^b}{d\tau} \, .
\ee 
Here $\tau$ is the euclidean time coordinate on $\mathbb{R}$ and the metric on $\cM_d$ can be expressed as
\be
g_{ab}(X) = \int_C \left( \frac{1}{e^2} \delta_a A \wedge * \delta_b A + \delta_a \phi \wedge * \delta_b  \bar \phi  + \delta_a \bar \phi  \wedge * \delta_b \phi \right)\, .
\ee

An electric impurity amounts to adding a term $(\sigma+iA_\tau)f$ to the original gauge theory lagrangian, where $f$ is an arbitrary function on $C$ and $\sigma$ is the scalar component of the vectormultiplet. It is shown in~\cite{Tong:2013iqa} that this results in a `dirty connection' $\tilde A_a (X)$ in the effective sigma-model action, 
\be
S = \int d\tau \ g_{ab} (X) \frac{dX^a}{d\tau} \frac{dX^b}{d\tau}  + \tilde{A}_a (X) \frac{dX^a}{d\tau} \, .
\ee
given by
\be
\tilde A_a (X) = \int_C    \alpha_{a} (X) \wedge * f\, .
\label{eq:dirtyconnection}
\ee

On the other hand, introducing a background line bundle $L_T$ for a topological flavour symmetry on $C$ amounts to adding the following contribution to the lagrangian of the supersymmetric gauge theory,
\be
\frac{1}{2\pi}(\sigma+iA_\tau) * F_T \, ,
\ee
where $F_T$ is the curvature of the connection on the holomorphic line bundle $L_T$. This is equivalent to an electric impurity with $f = \frac{1}{2\pi}* F_T$. The corresponding dirty connection is
\be
\widetilde A_{T,a} (X)= \frac{1}{2\pi}\int_{C }F_T \wedge \al_a(X) \, ,
\ee
which can be written more invariantly using the universal connection~\eqref{eq:univ-conn} as follows
\be
\widetilde A_T(X) = \frac{1}{2\pi}\int_{C }F_T \wedge \mathcal{A}(X) \, .
\ee
The curvature of the dirty connection is
\be
F_{\widetilde A_T} (X) = \frac{1}{2\pi}\int_{C} F_T \wedge  F_\mathcal{A}(X) \, .
\ee
We therefore see that the curvature of the dirty connection agrees with the curvature of the holomorphic line bundle $\widetilde L_T$ constructed using Deligne pairing in equation~\eqref{eq:DeligneCurvature}.

\section{Tangent Spaces from Massless Fermions}

\label{app:tangent}

In section~\ref{sec:U(1)1/2}, we stated that the contribution to the space of supersymmetric vacua with flux $\fm$ is captured by a supersymmetric sigma model to the moduli space $\cM_\fm = \sym^{\fm_\Phi} (C)$, where $\fm_\Phi = \fm+g-1 >0$. An important consistency check is that the massless Fermion fluctuations transform in the tangent space of $\mathcal{M}_\fm$. 

Let us first recall the construction of the tangent space to $\cM_\fm = \sym^{\fm_\Phi} (C)$. From the perspective of parametrizing effective divisors of degree $\fm_\Phi$, the tangent space at a divisor $D$ is given by
\be
T_D\cM_\fm = H^0(C,\cO (D) / \cO) \, .
\label{eq:sym-tangent}
\ee
This has a simple explanation when the divisor $D=p_1 + \cdots + p_{\fm_\Phi}$ consists of separated points and therefore
\be
T_D  \cM_\fm = \times_{i=1}^{\fm_\Phi} T_{p_i}C \, .
\ee
Each cotangent space $T^*_pC$ can be identified with $\fm_p /(\fm_p)^2$, where $\fm_p$ is the ideal of holomorphic functions vanishing at $p$. From this point of view, the dual of $\fm_p /(\fm_p)^2$ is then the space of residues of meromorphic functions with a simple pole at $p$, since there is a pairing given by multiplication and evaluation. These residues are exactly parametrised by~\eqref{eq:sym-tangent}.

Let us now write $L_\Phi$ for the holomorphic line bundle induced by $O(D)$. From the short exact sequence
\be
 0 \longrightarrow \cO \longrightarrow L_\Phi \longrightarrow L_\Phi / \cO \longrightarrow 0 \, ,
\ee
we see that the tangent space~\eqref{eq:sym-tangent} fits into a long exact sequence
\be
 0 \longrightarrow H^0(\cO) \overset{\alpha}{\longrightarrow} H^0(L_\Phi) \overset{\beta}{\longrightarrow} H^0(L_\Phi / \cO) \overset{\gamma}{\longrightarrow} H^1(\cO) \overset{\delta}{\longrightarrow} H^1(L_\Phi) \longrightarrow 0 \, .
 \label{eq:long-exact}
\ee
Notice that both maps $\alpha$ and $\delta$ are inherited from multiplication by the holomorphic section $\phi(D)$ whose zeros are parametrized by the divisor $D$. For fixed $D$, equation~\eqref{eq:long-exact} is a long exact sequence of vector spaces, which splits. In particular, 
\be
H^0(L_\Phi) = \mathrm{im}(\beta) \oplus \mathrm{coker}(\beta) \, ,
\ee
and so we can reconstruct the tangent space $T_D \cM_\fm$ from $\mathrm{im}(\beta)$ and $\mathrm{coker}(\beta)$ separately. This means that we can consider the following two short exact sequences
\bea
 &0 \longrightarrow H^0(\cO) \overset{\alpha}{\longrightarrow} H^0(L_\Phi) \overset{\beta}{\longrightarrow}  \mathrm{im}(\beta) \longrightarrow 0 \\
 &0 \longrightarrow \mathrm{coker}(\beta) \overset{\gamma}{\longrightarrow} H^1(\cO) \overset{\delta}{\longrightarrow} H^1(L_\Phi) \longrightarrow 0 \, .
\label{eq:short-exact-split}
\eea
The first line is the Euler sequence for
\be
j^{-1}(L_\Phi) = \mathbb{P}H^{0}(L_\Phi) \, ,
\ee
where $j: \sym^{\fm_\Phi} (C)\rightarrow J_C$ is the holomorphic map to the Jacobian parametrizing $L_\Phi$. Furthermore, in the case $\fm_{\Phi} > 2g-2$, $H^1(L_\Phi) = 0$ and so $\mathrm{coker}(\beta)=H^1(\cO)$. The map $\gamma$ becomes surjective and corresponds to the derivative of the projection $j$.

We are now in a position to explain how each summand $\mathrm{ker}(\beta)$ and $\mathrm{coker}(\beta)$ arises from the massless fermionic fluctuations by analysing Yukawa couplings in the original 3d supersymmetric gauge theory. In order to recover $\mathrm{ker}(\beta)$, we first note that terms in the first short exact sequence in~\eqref{eq:short-exact-split} correspond to the fermions:
\begin{itemize}
\item $H^0(L_\Phi)$: fermions $\psi$ in the $\cN=(0,2)$ chiral multiplet obtained from decomposition of the 3d chiral multiplet $\Phi$.
\item $H^0(\cO)$: gauginos $\Lambda_0$ in the $\cN=(0,2)$ vectormultiplet.
\end{itemize}
In the supersymmetric quantum mechanics, there is a Yukawa coupling proportional to
\be
 \int_C \bar\phi  \Lambda_0 \wedge * \psi \, .
\ee
We now note that since the map $\alpha: H^0(\cO) \to H^0(L_\Phi)$ corresponds to multiplication by $\phi$, a Fermion element in the image of $\alpha$ has the form $\psi = \phi \psi'$ where $\psi' \in H^0(\cO)$. On the image of $\al$, the Yukawa coupling then becomes
\be
 \psi' \Lambda_0\int_C * | \phi|^2,
\ee
which (provided $\zeta > \frac{2\pi \fm}{e^2\mathrm{Vol}_C}$ so that the moduli space is non-empty) generates a mass term for $\psi'$. As a consequence, we find that
\be
\mathrm{im}(\beta) = H^0(L_\Phi)/H^0(\cO)
\ee
parametrises the remaining massless fluctuations.

The second short exact sequence in~\eqref{eq:short-exact-split} arises from the remaining fermions:
\begin{itemize}
\item $H^1(L_\Phi)$ : fermions $\eta$ in the $\cN=(0,2)$ Fermi multiplet obtained from decomposition of the 3d chiral multiplet $\Phi$.
\item $H^1(\cO)$ : gauginos $\bar \Lambda$ in the $\cN=(0,2)$ chiral multiplet containing the covariant derivative $\bar D$.
\end{itemize}
In the supersymmetric quantum mechanics, there is a Yukawa coupling proportional to
\be
\int_C  \bar\Lambda \wedge * \langle \phi , \bar\eta\rangle \, .
\label{eq:Yukawa2}
\ee
The second short exact sequence~\eqref{eq:short-exact-split}
implies that the image of $\mathrm{coker}(\beta)$ in $H^1(\cO)$ is $\mathrm{ker}(\delta)$. Since the map $\mathrm{\delta} : H^1(\cO) \to H^1(L_\Phi)$ corresponds to multiplication by the holomorphic section $\phi$, a Fermion fluctuation $\bar \Lambda$ is in its kernel if and only if the product $\bar \Lambda \phi$ vanishes in the cohomology $H^1(L_{\Phi})$. Equivalently, $\bar \Lambda \phi = \bar \partial \lambda$ for some $\lambda \in H^0(L_{\Phi})$. This is the case if and only if the Yukawa coupling~\eqref{eq:Yukawa2} vanishes for each $\bar \eta$ and the Fermion remains massless.

\bibliographystyle{JHEP}
\bibliography{hilb}

\end{document}